\documentclass[a4paper,UKenglish,cleveref, autoref, thm-restate]{lipics-v2021}

\nolinenumbers 

\usepackage{wrapfig}
\usepackage{listings}
\usepackage{xcolor}
\usepackage{mdframed}
\usepackage{booktabs}
\usepackage{cite}

\newmdenv[
    backgroundcolor=yellow!30, 
    linewidth=0pt, 
]{yellowbackground}

\definecolor{mygray}{rgb}{0.92,0.92,0.92}

\lstdefinelanguage{CCustom}{
  language=C,
  morekeywords={uint64_t,unit},
}

\lstdefinelanguage{PyCustom}{
  language=Python,
  morekeywords={case, match}
}

\lstdefinelanguage{sail}{
  morekeywords={function, clause, union, enum, mapping, let, match, val,
  scattered, forall},
  morecomment=[l]{//}
}

\lstdefinelanguage{jibcustom}{
  alsoletter={\%},
  keywordsprefix={\%},
  morekeywords=[2]{fn, end, jump, goto},
  morecomment=[l]{//},
  keywordstyle=[2]\color{red},
  keywordstyle=[1]\color{teal},
}

\lstdefinelanguage{trace}{
    morecomment=[l]\#,
}

\lstdefinelanguage{opt}{
    morecomment=[l]\#,
}

\title{Pydrofoil: accelerating Sail-based instruction set simulators} \titlerunning{Pydrofoil}

\author
    {Carl Friedrich {Bolz-Tereick}}
    {Heinrich-Heine-Universität Düsseldorf, Germany \and \url{https://cfbolz.de/}}
    {cfbolz@gmx.de}
    {https://orcid.org/0000-0003-4562-1356}
    {}

\author
    {Luke {Panayi}}
    {Imperial College London, UK \and \url{https://wp.doc.ic.ac.uk/spo/person/291-2/}}
    {l.panayi21@imperial.ac.uk}
    {}
    {}

\author
    {Ferdia {McKeogh}} 
    {University of St Andrews, UK \and \url{https://github.com/fmckeogh}}  
    {fm208@st-andrews.ac.uk}
    {https://orcid.org/0009-0006-2772-0448}
    {}    
    
\author
    {Tom {Spink}}
    {University of St Andrews, UK \and \url{https://tom-spink.com/}}
    {tcs6@st-andrews.ac.uk}
    {https://orcid.org/0000-0002-7662-3146}
    {}

\author
    {Martin {Berger}}
    {University of Sussex, UK \and Montanarius Ltd, UK \and \url{https://martinfriedrichberger.net/}}
    {contact@martinfriedrichberger.net}
    {https://orcid.org/0000-0003-3239-5812}
    {}

\authorrunning{C.~F.~Bolz-Tereick et al.}


\Copyright{Carl Friedrich Bolz-Tereick and Luke Panayi and Ferdia McKeogh and Tom Spink and Martin Berger} 


\relatedversion{Full version at \url{https://arxiv.org/abs/TBC}}

\supplement{Artifacts for Reproducibility at \url{github...}}

\acknowledgements{We thank
  Alasdair Armstrong,
  Max Bernstein,
  Tim Hutt,
  Yutaka Nagashima,
  Amir Naseredini,
  Alastair Reid,
  Niyas Sait,
  Laurie Tratt,
  Mojtaba Valizadeh
  and the anonymous reviewers for their helpful comments.
  We thank Nico Rittinghaus for implementing Tristate numbers in the RPython JIT.}

\newcommand{\NI}{\noindent}
\newcommand{\EG}{e.g.,\ } 
\newenvironment{FIGURE}{\begin{figure}\rule{\linewidth}{.5pt}\centering}{\rule{\linewidth}{.5pt}\end{figure}}
\newenvironment{FIGURETOP}{\begin{figure}[t]\rule{\linewidth}{.5pt}\centering}{\rule{\linewidth}{.5pt}\end{figure}}
\newcommand{\IMAGESIZE}[2]{\begin{center}\includegraphics[width=#1\textwidth]{images/#2}\end{center}}
\newcommand{\IMAGE}[1]{\begin{center}\includegraphics[width=0.8\textwidth]{images/#1}\end{center}}
\newcommand{\EMPH}[1]{\emph{#1}}
\usepackage{ifthen}
\newboolean{showcomments}
\setboolean{showcomments}{true}
\ifthenelse{\boolean{showcomments}}
  {\newcommand{\mynote}[2]{
    \fbox{\bfseries\sffamily\scriptsize#1}
    {\small$\blacktriangleright$\textsf{\emph{#2}}$\blacktriangleleft$}
   }
  }
  {\newcommand{\mynote}[2]{}
  }

\newcommand{\CODE}[1]{\colorbox{mygray}{$\mathtt{#1}$}}
\newcommand{\NOVSPACEPARAGRAPH}[1]{\NI\textbf{\emph{#1}.}}
\newcommand{\PARAGRAPH}[1]{\vspace{2mm}\NOVSPACEPARAGRAPH{#1}}
\newcommand{\ANONYMISED}[1]{#1
}

\newcommand{\qemuspeedup}{26.7$\times$ }

\EventEditors{John Q. Open and Joan R. Access}
\EventNoEds{2}
\EventLongTitle{42nd Conference on Very Important Topics (CVIT 2016)}
\EventLongTitle{TBC}
\EventShortTitle{TBC 2025}
\EventAcronym{TBC}
\EventYear{2025}
\EventDate{TBC}
\EventLocation{TBC}
\EventLogo{}
\SeriesVolume{42}
\ArticleNo{23}

\begin{document}

\maketitle

\begin{abstract}

  \NI We present \EMPH{Pydrofoil}, a multi-stage compiler that
  generates instruction set simulators (ISSs) from processor
  instruction set architectures (ISAs) expressed in the high-level,
  verification-oriented ISA specification language Sail. Pydrofoil
  shows a > 230$\times$ speedup over the C-based ISS generated
  by Sail on our benchmarks, and is based on the following insights.
  (i) An ISS is effectively an interpreter loop, and tracing
  just-in-time (JIT) compilers have proven effective at accelerating
  those, albeit mostly for dynamically typed languages.  (ii) ISS
  workloads are highly atypical, dominated by intensive bit
  manipulation operations. Conventional compiler optimisations for
  general-purpose programming languages have limited impact for
  speeding up such workloads. We develop suitable domain-specific
  optimisations.  (iii) Neither tracing JIT compilers, nor
  ahead-of-time (AOT) compilation alone, even with domain-specific
  optimisations, suffice for the generation of performant ISSs.
  Pydrofoil therefore implements a
  hybrid approach, pairing an AOT compiler with a tracing JIT built on
  the meta-tracing PyPy framework.  AOT and JIT use domain-specific
  optimisations. Our benchmarks demonstrate that
  combining AOT and JIT compilers provides significantly greater
  performance gains than using either compiler alone.

\end{abstract}

\section{Introduction}\label{section_introduction}

A processor's functionality is defined by its \EMPH{instruction set
architecture} (ISA), which specifies the machine language interpreted
by the hardware. Accurate and comprehensive ISA specifications are
essential for processor design, compiler development, and
verification. \EMPH{Instruction set simulators} (ISSs), also called functional models,
provide software-based representations of processor behaviour.
The efficiency of ISS execution directly impacts processor development speed and associated costs.
Throughout this paper,
unless otherwise noted, ISS refers to full system simulation,
encompassing the entire hardware and software stack, not user-mode
simulation, which models only the application layer and sends system calls to the host's operating system
layer~\cite{RandalA:idevertrrthovmac,QEMU:sysemu}. 

Traditionally, ISSs were manually implemented, a process that becomes
increasingly time-consuming and error-prone as ISA complexity grows. To
address this scalability challenge, \EMPH{architecture definition
  languages} (ADLs) emerged, allowing for automated ISS generation
from formal ISA descriptions~\cite{MishraP:prodesl}. These ADLs aim to
streamline the processor development process. Performance overhead
associated with generated ISSs is often mitigated through just-in-time (JIT) compilation or partial
evaluation.  While ISSs are crucial for functional verification,
computer architects also require cycle-accurate models and hardware
synthesis capabilities. Consequently, many widely used ADLs, such as
SystemC~\cite{IEEE:ieestafsslrm}, extend beyond pure ISA specification
to incorporate microarchitectural details, including pipelining
information. This integration, while providing comprehensive modelling
capabilities, typically uses low-level language constructs, such as
C++ classes and macros, which introduce complexity into the
specification and verification process.

Sail~\cite{ArmstrongA:isasemfaracm} is an alternative to low-level
ADLs and eschews all hardware implementation detail in favour of
ISA-only modelling.  Sail streamlines ISA specification with high-level language
features such as arbitrary-precision arithmetic, pattern matching, and
type-based bitwidth-generics. Sail supports formal verification by also compiling
to logic, targeting theorem provers like Isabelle/HOL, Lean, and Z3.
Notably, Sail has been adopted as the official ISA-specification language
for RISC-V~\cite{riscvi:isaforsprp}.  However, the Sail-generated ISS
suffers from performance limitations. This leads to our
research agenda: \EMPH{how can a high-level ISA-specification
  language like Sail be compiled into performant
  ISSs?} The unique combination of Sail's high-level abstraction and
focus on verification challenges the applicability of conventional ISS
compilation techniques, including existing JIT-based approaches.
Concretely, our paper seeks to answer the following research
questions:
\begin{itemize}

\item \textbf{RQ1.} What are the bottlenecks causing the
  Sail-generated ISSs to be slow?

\item \textbf{RQ2.} Can tracing JIT compilation be used to speed
  up Sail-generated ISSs significantly?

\item \textbf{RQ3.} How close can we get to the performance of an
  industrial strength ISS, hand-tuned to RISC-V?

\end{itemize}

\PARAGRAPH{Contributions}
To address the research questions, we present \EMPH{Pydrofoil}, an
open-source\footnote{\ANONYMISED{\url{https://github.com/pydrofoil/pydrofoil/}}}
tracing JIT compiler for Sail, implemented using the RPython/PyPy
meta-tracing framework\footnote{\url{https://pypy.org/}}. We developed
and evaluated a suite of domain-specific optimisations tailored to the
unique characteristics of ISS workloads, which differ
from the typical Python applications targeted by PyPy. Our evaluation
methodology employs established computer architecture benchmarks,
primarily Linux-boot and SPEC workloads, alongside detailed ablation
studies to quantify the impact of individual optimisations.  In
evaluations using the RISC-V Sail model, Pydrofoil exhibited a >230$\times$
speedup over the Sail-generated ISS,  while exhibiting a \qemuspeedup
slowdown compared to handwritten and optimised QEMU.  This achievement shows
that whilst we make significant improvements over the Sail interpreter, there
is still performance to be gained.  We discuss methods of achieving this in
\autoref{sec-future}.

\section{Sail and the RISC-V Sail model}\label{section_background}

\subsection{Sail}

Sail is a domain-specific language (DSL) within the ML family~\cite{GordonM:metforipil,MilnerR:defstamlr},
designed to facilitate
processor specification and verification.  Sail hybridises first class
support for stateful programming with a rich typing system, including
algebraic-, \mbox{singleton-,} dependent- and higher-kinded types, 
as well as domain-specific arbitrary-precision and bitvector types, and a register construct.
The compilation pipeline of Sail is depicted in
Figure~\ref{image_sail_compilation_pipeline}. Beyond its primary
objective of enabling processor specification and verification through theorem proving,
Sail's most noteworthy changes from other ML-family languages are:
\begin{itemize}

\item Sail restricts function definitions to first-order, eliminating
  higher-order functions.

\item Sail supports \EMPH{liquid types}~\cite{RondonPM:liqtyp,XiH:depmlaatppwdt}.

\end{itemize}

\NI Liquid types, a form of dependent types, strike a balance between
expressive power and decidable type inference. They are useful for
addressing the bitwidth parameterisation inherent in modern processor
architectures. For instance, the RISC-V ISA supports both 32-bit (RV32)
and 64-bit (RV64) variants, where
the majority of the ISA specification remains parametric with respect
to the bitwidth, represented by a type declaration \CODE{xlen}.
Traditional ADLs often lack strong typing system support for bitwidth-generic data structures,
instead resorting to compile-time meta-programming techniques such as
the C preprocessor or C++ templates. Meta-programming
simplifies compiler optimisations, but necessitates redundant
verification across different bitwidth
configurations~\cite{BergerTrattLMCS}. We will argue in Section
\ref{sec-future} that the current Sail compiler may not
fully exploit the optimisation potential afforded by liquid type usage
within the RISC-V specification.

This paper primarily focuses on
domain-specific compiler design rather than programming language
ergonomics. Consequently, a detailed exposition of Sail's language
features is beyond the scope of this work. However, it is noteworthy
that Sail's capacity to seamlessly support both simulation and
logic-based verification contributed to its adoption as the official
specification language for RISC-V~\cite{riscvi:isaforsprp}.

\begin{FIGURE}
  \IMAGE{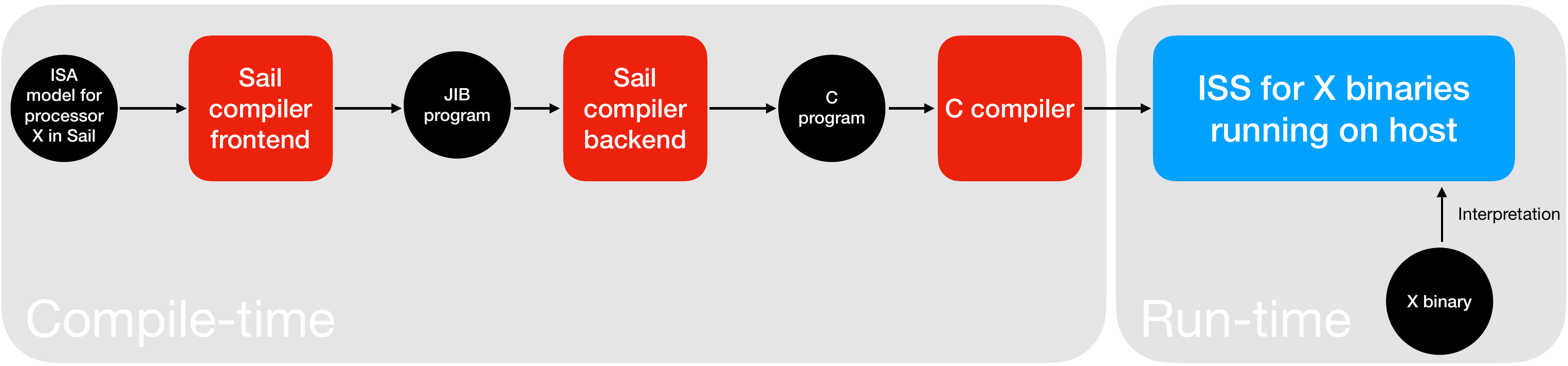}
  \caption{Sail's multi-stage compilation pipeline from Sail to an ISS running as a static binary.}
  \label{image_sail_compilation_pipeline}
\end{FIGURE}

\subsection{The RISC-V instruction set architecture}

RISC-V~\cite{ArsanovicK:inssetsbftcfrv,WatermanAdesrisvisa} is an
open-source ISA designed to be modular. Unlike proprietary ISAs
such as x86, Arm, and MIPS, RISC-V is vendor-neutral. This open nature
facilitates academic research and industrial adoption.  RISC-V
``\EMPH{is structured as a small base ISA with a variety of optional
  extensions. The base ISA is simple. [...] The optional extensions
  form a more powerful ISA for general purpose and high-performance
  computing.}''~\cite{WatermanAdesrisvisa}. This modularity, combined
with the availability of CHERI-RISC-V~\cite{watson:cheriisav9} and its
32-bit and 64-bit variants, results in a significant number of RISC-V
ISA configurations, rendering manual ISS development impractical.
Furthermore, the existence of RV32 and RV64 suggests the use of
bitwidth-generic data structures, particularly bitvectors. While
bitwidth-generics offer flexibility, they introduce compilation
complexities and potential performance overhead in ISS
implementations, as discussed in subsequent sections.

\subsection{The RISC-V model in Sail}\label{section_background_riscv_sail_model}

We shall now look at the a concrete ISA specification in Sail, in
order to illustrate how Sail supports the definition of ISA families,
using the official RISC-V formal specification~\cite{riscvigoldenmodel} as running example.

Abstractly, the toplevel function of the RISC-V Sail model is a big
loop, called \EMPH{fetch-decode-execute loop} in computer
architecture. In each round, the loop fetches the next
instruction from the external main memory (which an ISS simulates) as
a bit pattern, then decodes that pattern into a concrete instruction,
and finally calls an \CODE{execute} function that executes the decoded
instruction. As a simplification, the following pseudo-code serves as a
useful mental model:

\begin{FIGURE}
 \IMAGE{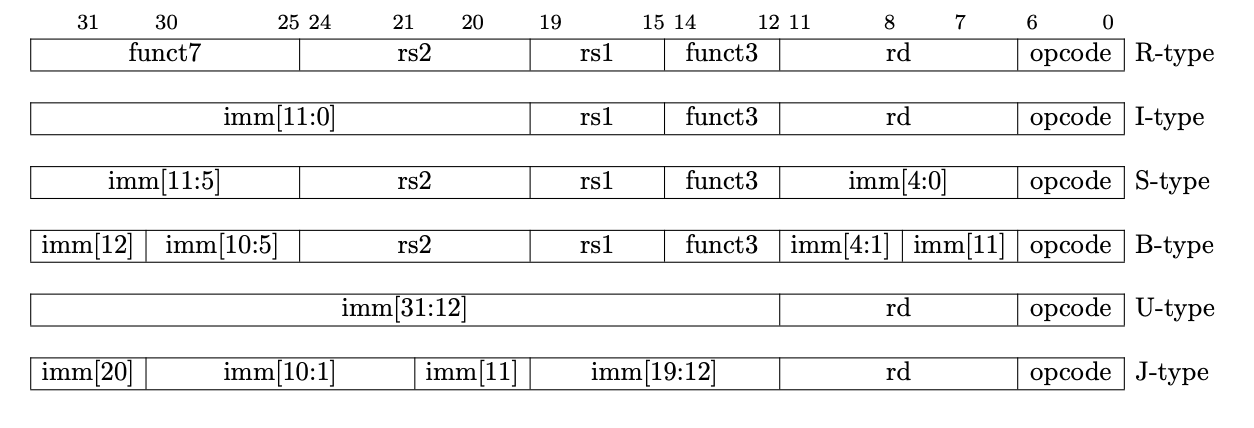}
 \caption{Two of the RISC-V instruction formats~\cite{WatermanA:riscvISAvolI}.}
  \label{figure_riscv-instruction-formats}
\end{FIGURE}

\begin{lstlisting}[language=PyCustom]
initialise_processor()
pc = 0
while true:
   cmd = memory[pc]
   pc += 4
   match decode(cmd):
      case addi(rd, rs1, imm):
         execute_addi(rd, rs1, imm)
      case xori(rd, rs1, imm):
         execute_xori(rd, rs1, imm)
      ...
      case _:
         ... # Illegal instruction
\end{lstlisting}

\NI Let us focus our attention on the addition
instruction, which, in RISC-V, has the mnemonic
\CODE{addi\ rd,\ rs1,\ imm}.
This instruction adds the 12-bit integer \CODE{imm} to the content of a
source register \CODE{rs1}.  The result is stored in the destination
register \CODE{rd}.  Like most ISAs, RISC-V uses instruction formats,
which can be seen as rough classifications of instructions that
aid decoding bit patterns into instructions. Figure
\ref{figure_riscv-instruction-formats} shows the structure of two
RISC-V instruction formats.  Our
example \CODE{addi\ rd,\ rs1,\ imm} is an instance of the \CODE{ITYPE}
format.  Like all \CODE{ITYPE} instructions,
\begin{itemize}

\item  \CODE{ITYPE} has opcode \CODE{0b0010011}, found in bits 0-6.

\item The destination register of the instruction is identified in bits 7-11,
  while the source register is given by bits 15-19.

\item The actual function to be computed is in the \CODE{funct3} field
  (bits 12 - 14): \CODE{addi} uses \CODE{0b000} here, while, \EG
  \CODE{0b100} identifies \CODE{xori}.

\item The 12-bit constant to be added to the source register is
   in bits 20-31.

\end{itemize}
Let us look at a simplified form of the Sail implementation of
\CODE{addi}.\footnote{See \url{https://github.com/riscv/sail-riscv/blob/2dfc4ff9f2bed3dcd0a3e8748211c99099e70ab7/model/riscv_insts_base.sail\#L159}~\cite{riscv_insts_base} for full code.}

\begin{lstlisting}[language=sail]
scattered union ast

val execute : ast -> Retired
scattered function execute

val encdec : ast <-> bits(32)
scattered mapping encdec

union clause ast = ILLEGAL : word
  
\end{lstlisting}

\NI First some type declarations: \CODE{ast}, short for
abstract syntax tree, is the disjoint union type, representing
all decoded instructions.  In Sail, \CODE{ast} is a \EMPH{scattered}
union, which means that the definitions of the variants are allowed to
be in different files.  This is useful when
processor ISAs evolve: we can add new instruction formats in new
files \EMPH{without} needing to modify existing code.
The function \CODE{execute} (also scattered) specifies the actual
semantics of RISC-V instructions, meaning the changes to processor state
induced by executing \CODE{addi}. The \CODE{execute} function takes an \CODE{ast} as argument
and returns \CODE{Retired}, which is an \CODE{enum} specifying whether
the instruction successfully executed or not (execution can fail,
for example due to interrupts). The RISC-V model is
imperative: \CODE{execute} can and does modify processor state, in
particular registers, when invoked. Such processor state is global.

To translate between the encoding of instructions in bits, and their
more structured representation as \CODE{ast}, the RISC-V model uses the
(scattered) mapping \CODE{encdec}, short for
``encoding-decoding''. The keyword \CODE{mapping} is for
\EMPH{bidirectional} function definitions~\cite{FosterJN:bidprol}. This means \CODE{encdec}
can either take an \CODE{ast} and encode it into a 32-bit bitvector, or
take a bitvector and decode it into an \CODE{ast}.  This reflects, and the
typing system enforces, the (almost) one-to-one relationship between ASTs and
bitvectors---almost, because not all 32-bit patterns correspond to
valid instructions. The model has  a fall-through clause
\CODE{union\ clause\ ast = ILLEGAL : word}.

We'll now look at the clauses of \CODE{ast}, \CODE{encdec} and
\CODE{execute} for the \CODE{ITYPE} format.

\begin{lstlisting}[language=sail]
enum iop = {RISCV_ADDI, RISCV_SLTI, RISCV_SLTIU, RISCV_XORI, RISCV_ORI, RISCV_ANDI}    (*@\label{code_generated_ITYPE_defs_1}@*)
  
...

union clause ast = ITYPE : (bits(12), regidx, regidx, iop) (*@\label{code_generated_ITYPE_defs_2}@*)

\end{lstlisting}

\NI Line \ref{code_generated_ITYPE_defs_1} enumerates all \CODE{ITYPE}
instructions and names this type \CODE{iop}.  Line
\ref{code_generated_ITYPE_defs_2}, containing
\CODE{union\ clause\ ast\ =\ ITYPE\ : ...}, defines the variant of the
\CODE{ast} type that holds \CODE{ITYPE} instructions. This clause
carries a tuple of information: \CODE{(bits(12),\ regidx,\ regidx,\ iop)}

\NI The first component holds 12 bits using the type \CODE{bits(12)}.
This is the constant to be added to the source register. The next two
components identify the source and target registers, where
\CODE{regidx} abbreviates the type \CODE{bits(5)}, the type of
bitvectors of length 5.  (RISC-V has 32 registers, so 5 bits suffice.)
We also need to translate between the 3-bit opcode field in the
\CODE{ITYPE} instruction format and the enumeration \CODE{iop}.  This
is done by the following bidirectional \CODE{mapping}:

\begin{lstlisting}[language=sail]
mapping encdec_iop : iop <-> bits(3) = {
  RISCV_ADDI  <-> 0b000,
  RISCV_SLTI  <-> 0b010,
  RISCV_SLTIU <-> 0b011,
  RISCV_ANDI  <-> 0b111,
  RISCV_ORI   <-> 0b110,
  RISCV_XORI  <-> 0b100
}

mapping clause encdec = ITYPE(imm, rs1, rd, op)
  <-> imm @ rs1 @ encdec_iop(op) @ rd @ 0b0010011
\end{lstlisting}

\NI The actual semantics of the \CODE{ITYPE} instructions is given by
the following \CODE{execute} clause:

\begin{lstlisting}[language=sail]
function clause execute (ITYPE (imm, rs1, rd, op)) = {               (*@\label{code_generated_TYPE-EXEC-1}@*)
  let rs1_val = X(rs1);                                              (*@\label{code_generated_TYPE-EXEC-2}@*)
  let immext : bits(64) = sign_extend(imm);                                     (*@\label{code_generated_TYPE-EXEC-3}@*)
  let result : bits(64) = match op {                                            (*@\label{code_generated_TYPE-EXEC-4}@*)
    RISCV_ADDI  => rs1_val + immext,                                 (*@\label{code_generated_TYPE-EXEC-5}@*)
    RISCV_SLTI  => zero_extend(bool_to_bits(rs1_val <_s immext)),
    RISCV_SLTIU => zero_extend(bool_to_bits(rs1_val <_u immext)),
    RISCV_ANDI  => rs1_val & immext,
    RISCV_ORI   => rs1_val | immext,
    RISCV_XORI  => rs1_val ^ immext
  };
  X(rd) = result;              (*@\label{code_generated_TYPE-EXEC-6}@*)
  RETIRE_SUCCESS               (*@\label{code_generated_TYPE-EXEC-7}@*)
}
  
\end{lstlisting}

\NI The code is beautifully simple and clear. However, the simplicity
hides the following complexities:

\begin{itemize}

\item The function definition clause in Line \ref{code_generated_TYPE-EXEC-1} pattern matches on the
  \CODE{ast} data type, triggers when the instruction is \CODE{ITYPE},  and gives convenient, locally-scoped names to
  the components \CODE{imm}, \CODE{rs1}, \CODE{rd} and \CODE{op},
  which were chosen to match the corresponding names in~\cite{WatermanA:riscvISAvolI}.

\item Line \ref{code_generated_TYPE-EXEC-2} loads the register value
  to be added from \CODE{rs1}, the source register.

\item Line \ref{code_generated_TYPE-EXEC-3} sign-extends the 12 bits
  value \CODE{imm} to the target register bitwidth (32 or 64 bits,
  depending on the version of RISC-V used).  Two things are noteworthy
  about \CODE{sign\_extend(imm)} and its use here.  First,
  \CODE{sign\_extend(imm)} is parameterised with source and
  target bitwidths, albeit subject to the requirement that the latter not be
  smaller than the former. This bitwidth parameterisation is visible in the
  function's type
  \begin{center}
  \CODE{forall\ 'n\ 'm,\ 'm\ \geq\ 'n.\ (implicit('m),\ bits('n)) \rightarrow bits('m)}
  \end{center}
  which has type parameters \CODE{'n} and \CODE{'m} that range over
  integers (singleton types). The predicate \CODE{'m\ \geq\ 'n} is a
  refinement predicate, evaluated during type checking.  We refer to
  the Sail manual~\cite{ArmstrongA:saiinssssl} for a detailed
  explanation of Sail's type syntax.  Second, while
  \CODE{sign\_extend(imm)} takes two arguments, in Line
  \ref{code_generated_TYPE-EXEC-3} it is called with only one: the
  required target bitwidth \CODE{'m} is not specified in the call.
  The keyword \CODE{implicit} gives it away: Sail has a restricted
  form of implicit function parameterisation~\cite{LewisJR:imppardswst,OderskyM:simfouaaoift}, albeit resolved at
  compile-time. The resolution of implicit arguments is restricted to
  the expected output type, here \CODE{bits(64)}.

\item Line \ref{code_generated_TYPE-EXEC-4} pattern matches on the
  exact function to be executed.

\item Line \ref{code_generated_TYPE-EXEC-5} triggers if we match
  \CODE{RISCV\_ADDI}. In this case, the integer \CODE{imm} encoded in
  the instruction and suitably sign-extended, is added to the content
  in the source register by \CODE{rs1\_val + immext}.

\item Line \ref{code_generated_TYPE-EXEC-6} writes the result into the
  destination register \CODE{rd}.

\item Finally, Line \ref{code_generated_TYPE-EXEC-7} returns an
  indication that the instruction has been successfully executed to
  the caller.

\end{itemize}
For this code, Sail offers strong \EMPH{compile}-time guarantees,
for example, we can neither assign bitvectors of
different bitwidths, nor access a vector out-of-bounds.
Section \ref{sec-future} discusses future work on better compiler
optimisations based on Sail's expressive types.

\subsection{How does the Sail compiler generate an ISS?}

Sail generates a C-based ISS.  Unfortunately, this ISS is slow: on a
2024 laptop (AMD Ryzen 7 PRO 7840U, with 32 GB main memory) it executes less
than a million RISC-V instructions per second; QEMU simulates a few hundred million instructions in the same time.  In this section we
identify the root-causes of Sail's poor performance.  We start with two
possibly unsurprising remarks to set the scene.
\begin{itemize}

\item Sail is a high-level language that emphasises precise
  expression of ISA semantics over concern for simulation speed at
  every step. High-level languages are harder to generate fast code
  from than low-level languages.

\item Development of optimising compilers is labour intensive and
  there is a direct correlation between the level of optimisation
  achievable and developer hours spent.  The Sail community chose to
  spend the scarce developer resources available in a university lab
  on verification-related parts of the toolchain.

\end{itemize}

\NI Let us look at concrete code that shows where Sail leaves
performance on the table.  Figure \ref{fig-itype-c} shows C generated from the
\CODE{execute} clause of \CODE{ITYPE} instructions, discussed
above. The code is lightly edited for
readability, with Sail code added as comments before the corresponding C code
lines.
\begin{FIGURE}
\begin{lstlisting}[language=CCustom]
  {
    uint64_t imm = ...;
    ...
    // let rs1_val = X(rs1);
    uint64_t rs1_val;
    rs1_val = rX_bits(rs1);
    // let immext : xlenbits = sign_extend(imm);
    uint64_t immext;
    {
      i_generic i_generic_64;
      CREATE(i_generic)(&i_generic_64);
      CONVERT_OF(i_generic, int64_t)(&i_generic_64, INT64_C(64));
      bv_generic imm_bv_generic;
      CREATE(bv_generic)(&imm_bv_generic);
      CONVERT_OF(bv_generic, fbits)(&imm_bv_generic, imm, UINT64_C(12) , true);
      bv_generic immext_bv_generic;
      CREATE(bv_generic)(&immext_bv_generic);
      sign_extend(&immext_bv_generic, i_generic_64, imm_bv_generic); (*@\label{code_generated_1_action2}@*)
      immext = CONVERT_OF(fbits, bv_generic)(immext_bv_generic, true);
      KILL(bv_generic)(&immext_bv_generic);
      KILL(bv_generic)(&imm_bv_generic);
      KILL(i_generic)(&i_generic_64);
    }
    uint64_t result;
    {
      __label__ case_8084, finish_match_8083;
      {
        if ((RISCV_ADDI != op)) goto case_8084;
	// rs1_val + immext
        result = rs1_val + immext; (*@\label{code_generated_1_action1}@*)
        goto finish_match_8083;
      }
    case_8084: ;
      ... // the other cases
    }
    {
      wX_bits(rd, result);
    }
    cbz31056 = RETIRE_SUCCESS;
    goto finish_match_8062;
  }
\end{lstlisting}
    \caption{Fragment of the \CODE{execute} clause of \CODE{ITYPE} instructions
    after compilation to C by Sail.}
    \label{fig-itype-c}
\end{FIGURE}
\NI The computation in Line \ref{code_generated_1_action1} of the
result in the \CODE{ADDI} case is mapped to an addition on the host
machine. Sign extending \CODE{imm}, the 12-bit immediate, is done
by calling \CODE{sign\_extend} in Line
\ref{code_generated_1_action2}.  The arguments of this call are of
type \CODE{bv\_generic}, which is a \EMPH{heap}-allocated data
structure storing bitvectors of arbitrary width.  Converting \CODE{imm}
to \CODE{bv\_generic}, requires a heap allocation as part of the
\CODE{CONVERT\_OF} macro; \CODE{i\_generic} is also heap allocated and
represents integers of arbitrary size.\footnote{Those types are called
\CODE{lbits} and \CODE{sail\_int} in the \EMPH{actual} C code that
Sail generates. We change their names for consistency
with Section \ref{section_implementation}, where we discuss the datatypes in more detail.}  In Sail, both
\CODE{bv\_generic} and \CODE{i\_generic} are implemented using the
\EMPH{GNU Multiple Precision Arithmetic library (GMP)}~\cite{GMP}
for arbitrary precision arithmetic, operating on signed integers,
rational numbers, and floating-point numbers.  We summarise the key
reasons why the Sail-generated simulator is slow:
\begin{itemize}

\item \textbf{Bitvectors.} Sail tries to infer the width of the
  bitvector variables in the program. If they are statically known to
  fit into an unsigned 64-bit integer, the C backend will map them to
  the \CODE{uint64\_t} type. However, many functions in a Sail spec
  are parametrised with bitwidths, for example \CODE{sign\_extend}. The
  bitvectors in these functions are implemented using
  \CODE{bv\_generic} and thus ultimately become GMP integers. This has a
  number of drawbacks: all operations on GMP-based bitvectors require
  costly heap allocations, because GMP integer types are always
  heap-allocated.

\item \textbf{Integers.} Sail's integer type has arbitrary
  precision. As with bitvectors, Sail tries to infer whether a given
  integer variables can only have values that fit into a signed 64-bit
  integer. When that is possible, the C backend will use
  \CODE{int64\_t}. However, where it is impossible for Sail to prove
  that the values fits into \CODE{int64\_t}, the C backend uses an
  arbitrary-precision representation from GMP which requires
  heap allocations on every integer operation.

\item \textbf{Interpretation overhead.} Finally, Sail generates an
  interpreter, which has to re-analyse the bits of the program being
  simulated again and again. Note that actual processors use caches to
  ameliorate this overhead, \EG with an instruction cache.

\end{itemize}

\NI This brings us to our key objective: \textbf{how can we improve simulation performance?} 

\subsection{RPython}

Pydrofoil is implemented using the PyPy framework~\cite{rigo_pypys_2006}.
RPython~\cite{AnconaD:rpystetrdastol} is PyPy's implementation language.
RPython (short for Restricted Python) is a statically typable subset of Python2. The language was developed
to aid the implementation of high-performance virtual machines for dynamic
languages. To this end, RPython contains a generic meta-tracing
just-in-time compilation infrastructure~\cite{BolzCF:trametlptjc} that
can be retargeted to different guest languages.
Meta-tracing works by tracing through the main interpreter loop (written in
the host language---here RPython) of the guest language
interpreter~\cite{BalaV:dyntradynos,SullivanGT:dynnatoi,MitchellJG:desconofaeips,FisherJA:optohmwabbb,FisherJA:traswchatfgmc}.
The tracing JIT traces many iterations of the main interpreter loop,
effectively unrolling it. This means that one trace contains several guest language
instructions, which can then be optimised together.
The trace is optimised, compiled to machine code, and then available to execute
the corresponding guest language program fragment more efficiently.  As ISA models, whether
written in Sail or not, are essentially interpreters for an ISA, it
seems reasonable to expect that the RPython meta-tracing JIT can speed
up the Sail interpreter loop, and compile guest machine instructions
to host machine instructions. This has been confirmed in the Pydgin
project~\cite{LockhartD:pydgenfissfsadwmtjc}, which implements user-mode ISSs
for subsets of Arm v6, MIPS and RISC-V by writing interpreters for them manually
in RPython.

\section{How is Pydrofoil implemented?}\label{section_implementation}

\begin{FIGURE}
  \IMAGE{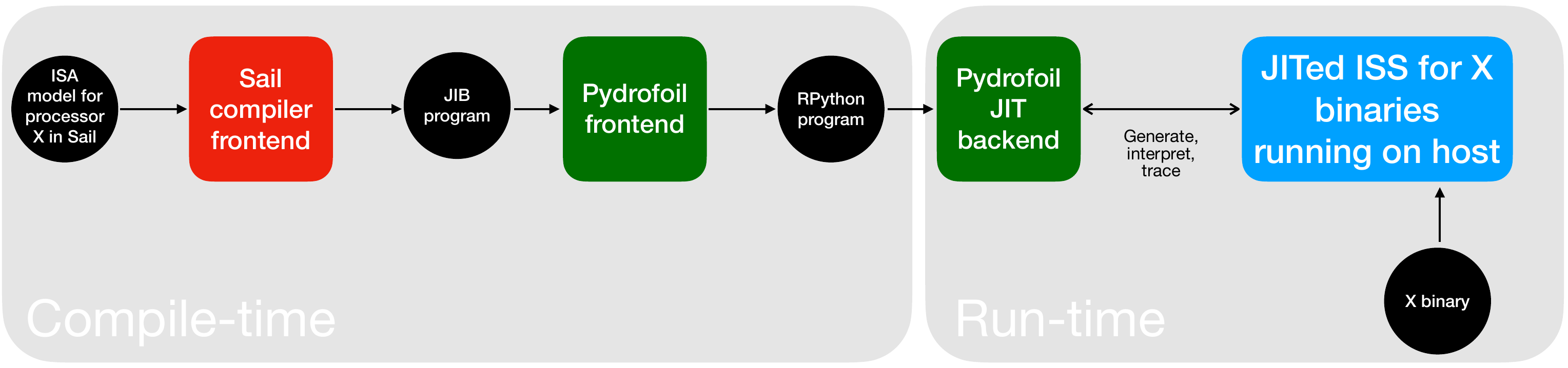}
  \caption{Pydrofoil multi-stage compilation pipeline from Sail to a JITed ISS.}
  \label{image_pydrofoil_compilation_pipeline}
\end{FIGURE}

Pydrofoil's approach for better ISS performance is based on the following
ideas:

\begin{itemize}

\item Better static optimisations of Sail code, in particular
  optimisations for bitvectors and integers, with the goal of minimising
  heap allocations.

\item Leveraging run-time information with the help of the RPython
  tracing JIT compiler for further optimisations not available
  statically.

\end{itemize}

\NI The first one is, at least in principle, also available to
Sail's C backend.

\subsection{Pydrofoil architecture}

The Sail system emits an ISS in C to be able to use
the existing ecosystem of optimising C compilers. In the same way, Pydrofoil reuses the Sail
frontend, RPython and the RPython JIT to generate a faster ISS
from a Sail model. Pydrofoil's overall architecture is shown in Figure
\ref{image_pydrofoil_compilation_pipeline}.

Pydrofoil's input is JIB, the Sail compiler's textual intermediate
representation~\cite{ArmstrongA:isla}.  When JIB is emitted, all of
Sail's type checking has been done, all variables have explicit types
and all pattern matching is compiled away into basic blocks and
(conditional) gotos. This makes JIB easy for the Pydrofoil
ahead-of-time (AOT) compiler to work with.  From JIB, Pydrofoil
generates its own SSA-based intermediate representation (IR)
for Sail functions. These functions are
optimised by Pydrofoil's static optimisation passes.  Then Pydrofoil
generates RPython source code from the optimised IR. The generated
code makes use of Pydrofoil's runtime library of support functions
that implement Sail datatypes, including problem cases like the
aforementioned bitvectors, and arbitrary precision integers.  The
support library also implements the simulated main memory that the
simulated processor interacts with.  This resulting RPython code is
saved to file, and can be executed with a normal Python2
implementation. This is not performant, but useful for testing.
Pydrofoil further translates it into a binary with the existing RPython
translation toolchain. During this translation, a tracing JIT compiler
can be (optionally) inserted into the resulting binary.  The following
sections will elaborate on key steps of this process.

\subsection{Static Pydrofoil IR optimisations}
\label{sec_static_opts}

Once JIB is translated to Pydrofoil's SSA-based IR, we apply a number of
standard compiler optimisations: constant folding, dead code elimination,
common subexpression elimination, inlining, and scalar replacement of
aggregates (using a simple form of partial escape analysis)~\cite{BolzCF:allrembpeiatj,Stadler:parescaasrfj}.  All static
optimisations are applied until a fixpoint is reached, and
further optimisation passes would not alter the IR.

\subsubsection{Static Optimisations of bitvector and integer operations}
\label{sec_static_opts_bv_int}

We focus our attention on static optimisations for operations on the
bitvector and integer data types, because of their importance for ISSs
in general, and Sail in particular.  JIB uses two different types for
each of these:\footnote{In line with the previous section, we renamed
two JIB types for consistency: \CODE{\%bv\_generic} is called \CODE{\%bv}
in JIB, whereas \CODE{\%i\_generic} is just called \CODE{\%i}.}

\begin{itemize}

  \item \textbf{Bitvectors:} if JIB can infer precise bitwidth, then
    bitvector types like \CODE{\%bv64} or \CODE{\%bv16} of fixed
    width are used. Otherwise \CODE{\%bv\_generic} is used, a generic bitvector type
    where the width is not known at runtime (which turns into the
    \CODE{bv\_generic}).

  \item \textbf{Integers:} if JIB can infer that an integer can always fit into
    a 64-bit signed machine integer then \CODE{\%i64} is used. Otherwise
    \CODE{\%i\_generic}, a type of arbitrary precision integer, is used.

\end{itemize}

\begin{FIGURE}
\begin{lstlisting}[language=jibcustom]
fn execute(mergez3var) {
...
  imm : %bv12
  imm = ...
  // let rs1_val = X(rs1);
  rs1_val : %bv64
  rs1_val = rX_bits(rs1)
  // let immext : xlenbits = sign_extend(imm); (*@\label{itype-execute_jib_1}@*)
  immext : %bv64                               (*@\label{itype-execute_jib_2}@*)
  sail_int_64 : %i_generic
  sail_int_64 = i64_to_i_generic(64) // cast   (*@\label{itype-execute_jib_cast_1}@*)
  imm_bv_generic : %bv_generic
  imm_bv_generic = imm               // cast   (*@\label{itype-execute_jib_cast_2}@*)
  immext_bv_generic : %bv_generic
  immext_bv_generic = sign_extend(sail_int_64, imm_bv_generic) (*@\label{itype-execute_jib_sign_extend}@*)
  immext = immext_bv_generic         // cast   (*@\label{itype-execute_jib_3}@*)
...
\end{lstlisting}
\caption{JIB code for the \CODE{execute} clause of \CODE{ITYPE} instructions.}
  \label{fig-itype-execute-jib}
\end{FIGURE}

\NI JIB code frequently casts between the generic and fixed-width
variants of the respective representations. Casts do not change the
underlying value, but cost a heap allocation when casting to a generic
datatype, and heap memory reads when casting in the other
direction. Both are expensive.  Another reason why the generic forms
\CODE{\%bv\_generic} and \CODE{\%i\_generic} are much less efficient
at runtime is that the operations (\EG addition for
\CODE{\%i\_generic}) cannot be mapped to directly to host CPU
operations, since we do not know their bitwidth. Here is a concrete
example. The JIB code in Figure~\ref{fig-itype-execute-jib}, lightly edited for
clarity, corresponds to the beginning of \CODE{ITYPE}'s \CODE{execute} fragment
that we already saw in Sail and C in Section
\ref{section_background_riscv_sail_model}, and Figure~\ref{fig-itype-c}.
Line~\ref{itype-execute_jib_1} in Figure \ref{fig-itype-execute-jib} shows the
\CODE{sign\_extend} from Sail in comments.  JIB Lines
\ref{itype-execute_jib_2}--\ref{itype-execute_jib_3}, execute this
\CODE{sign\_extend}: first the immediate value of type \CODE{\%bv12} is
cast to a generic bitvector (Line~\ref{itype-execute_jib_cast_2}), then
sign-extended to 64 bits (Line~\ref{itype-execute_jib_sign_extend}), and finally cast back to
\CODE{\%bv64} (Line~\ref{itype-execute_jib_3}). This sign extension alone by default requires three
heap allocations. The reason why these
casts are necessary is that the \CODE{sign\_extend} function takes a
\CODE{bv\_generic} and an \CODE{i\_generic} and returns a
\CODE{bv\_generic}.\footnote{Note that the JIB translation of \CODE{sign\_extend} takes
a bitvector and then an integer as input. We saw in Section
\ref{section_background_riscv_sail_model} that the order is reversed
in the eponymous Sail function. This is an artefact of Sail's
resolution of implicit arguments, and irrelevant for compiler
optimisations. It's implemented in the Sail prelude of the RISC-V model by
having a small one-operation Sail helper function that has the implicit
argument, which calls the JIB \CODE{sign\_extend} operation with arguments
reversed.}

For all built-in Sail functions and operations that deal with bitvectors
and integers, Pydrofoil implements a bundle of variants. They all perform the
same computation, but take more specific types as arguments.
This enables Pydrofoil to avoid some heap allocations, as we discuss next.
To optimise such operations on generic bitvectors and integers, Pydrofoil's
static optimiser uses peephole rewrite rules that inspect the IR of a
function and rewrite the operations on these types into more efficient
forms.  In the best case rewrites remove casts between generic and
specific bitvector types (as well as machine integers and arbitrary
integers) and use specialised operations on the types (when they are
available in Pydrofoil's runtime library).

Here are Pydrofoil's variants of the \CODE{sign\_extend} function,
together with the rewrite rules that the static optimiser uses:

\begin{lstlisting}[language=opt]
# functions

# nothing is known statically
sign_extend:             (%bv_generic, %i_generic) -> %bv_generic
# the target width is known to fit into an i64
sign_extend_g_i:         (%bv_generic, %i64) -> %bv_generic
# the target width is a constant
sign_extend_g_c<n>:      (%bv_generic) -> %bv<n>
# the argument bitvector has a constant width
sign_extend_bv_c<n, m>:  (%bv<n>) -> %bv<m>

# rewrite rules
sign_extend(b, i64_to_i_generic(i))             -> sign_extend_g_i(b, i)
sign_extend_g_i(b, c)                           -> sign_extend_g_c<c>(b) [if c <= 64]
sign_extend_g_c<c>(cast(b:%bv<n>, %bv_generic)) -> sign_extend_bv_c<n, c>(b, n, c)
\end{lstlisting}

\NI After applying the rewrite rules to the \CODE{sign\_extend} call in the
example code on Line~\ref{itype-execute_jib_sign_extend} in
Figure~\ref{fig-itype-execute-jib}, it can be replaced by a call to
\CODE{sign\_extend\_bv\_c\langle 12, 64 \rangle}. The latter takes a
\CODE{\%bv12} as argument, and returns a \CODE{\%bv64}. This means we get rid
of all three heap allocations.
The resulting code is equivalent to the following JIB code and
does not contain casts:

\begin{lstlisting}[language=jibcustom]
fn execute(mergez3var) {
...
  imm : %bv12
  imm = ...
  // let rs1_val = X(rs1);
  rs1_val : %bv64
  rs1_val = rX_bits(rs1)
  // let immext : xlenbits = sign_extend(imm);
  immext : %bv64
  immext = sign_extend_bv_c<12, 64>(imm)
...
\end{lstlisting}

To be able to narrow the generic integer type \CODE{\%i\_generic} to
\CODE{\%i64}, Pydrofoil also performs a range
analysis. If the analysis finds an integer variable where the range
fits into a signed 64-bit int, Pydrofoil narrows the type to
\CODE{\%i64}. An example is the RISC-V \CODE{mulw} instruction, which
multiplies two signed 32-bit bitvectors by converting them to
\CODE{\%i\_generic} first. The result of the multiplication of these two
integers must fit into an \CODE{\%i64}, so Pydrofoil narrows the integer types
and rewrites the multiplication operation accordingly.

\subsubsection{Interaction with inlining}
\label{sec_inlining}

Inlining helps to make bitvector and integer
operation rewrites more effective. Many Sail functions are small
helper functions that operate on generic bitvectors and integers,
passed as arguments. After inlining such small functions into their
use contexts, the concrete bitvector widths of the argument can often
be inferred from the calling context. Then the bitvector operations in
the inlined function are specialised to smaller bitwidths.

Let us look at another  \CODE{ITYPE} operation as an
example: the \CODE{slti} case uses the \CODE{operator\ <\_s},
to perform the signed comparison. It is defined as follows in Sail:

\begin{lstlisting}[language=sail]
val operator <_s  : forall 'n, 'n > 0. (bits('n), bits('n)) -> bool
  
function operator <_s  (x, y) = signed(x) < signed(y)
\end{lstlisting}

\NI Here \CODE{signed} is a Sail function that takes a bitvector of width
$n$, and returns the corresponding signed integer.  This code is
compiled to the JIB code in the top part of Figure~\ref{fig-operator}.
It implements a signed comparison of bitvectors and operates on generic
bitvectors. In the concrete context where the function is called, the bitvector
widths are mostly known. For example, its use in the \CODE{slti}
instruction can be seen in the bottom half of Figure~\ref{fig-operator}.

\begin{FIGURE}
\begin{lstlisting}[language=jibcustom]
val (operator <_s): (%bv_generic, %bv_generic) ->  %bool

fn (operator <_s)(x, y) {
  // signed(x) < signed(y)
  z40 : %i_generic
  z40 = signed(x)
  z41 : %i_generic
  z41 = signed(y)
  return = lt_int(z40, z41)
  end;
}
\end{lstlisting}
\begin{lstlisting}[language=jibcustom]
  jump @neq(RISCV_SLTI, z0) goto 280
  // RISCV_SLTI  => zero_extend(bool_to_bits(rs1_val <_s immext)),
  z2 : %bv1
  z6 : %bool
  z7 : %bv_generic
  z7 = rs1_val      // cast (*@\label{slti-jib-cast-1}@*)
  z8 : %bv_generic
  z8 = immext       // cast (*@\label{slti-jib-cast-2}@*)
  z6 = (operator <_s)(z7, z8)
  z2 = bool_to_bits(z6)
  ...
\end{lstlisting}
    \caption{JIB code for \CODE{operator\ <\_s} (top half) and its used in the
    JIB code of the \CODE{execute} clause of the \CODE{slti} instruction
    (bottom half).}
\label{fig-operator}
\end{FIGURE}

The arguments to  \CODE{operator\ <\_s} are cast from \CODE{\%bv64} to
\CODE{\%bv\_generic} (Lines \ref{slti-jib-cast-1} and \ref{slti-jib-cast-2}).
After inlining the call to \CODE{operator\ <\_s}, the calls to \CODE{signed} in its body can
therefore be optimised to a variant that takes a \CODE{\%bv64} and
returns an \CODE{\%i64}, which then allows the \CODE{lt\_int}
operation to be replaced by a version that operates on
\CODE{\%i64}. This removes \EMPH{all} heap allocations from
the inlined body of \CODE{operator\ <\_s} that replaces the call-site.


\subsubsection{Function specialisation}
\label{sec_specialisation}

In cases where a function is too large\footnote{We use an inlining limit of 25
JIB operations in up to 4 basic blocks at the moment.} to be inlined, Pydrofoil
tries to specialise it to the specific bitvector widths and \CODE{\%i64} types of its arguments at the function's call-sites.
The specialiser will go over all the non-inlinable function
calls in the program and check whether any of the arguments are of type
\CODE{\%bv\_generic} or \CODE{\%i\_generic} and either constant, or the result
of casts from more specific types. For such
calls, the target function will be copied into a more specific
version. The copied function takes the more specific argument types. The body of
the copy is then optimised to take the newly available bitvector widths of its
arguments into account. Sometimes, as the result of specialisation, the
return type of a function can also become more specific. This allows
the caller of such a specialised function to be optimised further.

\subsubsection{When do the static bitvector and integer optimisations fail?}\label{sec_implementation_when_fail}

Peephole rewrites for bitvector and integer operations work only
within single functions. They can only be applied when the arguments
of such operations are cast from more specific types in the same
function where the operation happens. Inlining and function
specialisation help with peephole rewrites being restricted to single
functions. But this is sometimes still insufficient for giving
peephole rewrites enough context to do their work. In more complicated
cases, particularly if the arguments of bitvector operations are read
out of a \CODE{union} or a \CODE{struct}, more powerful static
analysis would be needed to be able to optimise the operations to more
specific variants.  An example is the \CODE{LOAD} constructor in the RISC-V
specification, which implements loading data from simulated main memory into a
register. The Sail code can be seen in Figure~\ref{fig-load}.

\begin{FIGURE}
\begin{lstlisting}[language=sail]
val mem_read      : forall 'n, 0 < 'n <= max_mem_access . (AccessType(ext_access_type), physaddr, int('n), bool, bool, bool)       -> MemoryOpResult(bits(8 * 'n))

enum word_width = {BYTE, HALF, WORD, DOUBLE}

union clause ast = LOAD : (bits(12), regidx, regidx, bool, word_width, bool, bool)

val extend_value : forall 'n, 0 < 'n <= xlen. (bool, bits('n)) -> xlenbits
function extend_value(is_unsigned, value) = if is_unsigned then zero_extend(value) else sign_extend(value)

mapping size_bytes : word_width <-> {1, 2, 4, 8} = {
  BYTE   <-> 1,
  HALF   <-> 2,
  WORD   <-> 4,
  DOUBLE <-> 8,
}

function clause execute (LOAD(imm, rs1, rd, is_unsigned, width, aq, rl)) = {
  let offset : xlenbits = sign_extend(imm);
  let width_bytes = size_bytes(width);
  let addr = X(rs1) + offset;
  // simplified, the real code deals with virtual memory and checks for alignment
  match mem_read(Read(Data), addr, width_bytes, aq, rl, false) {
    Ok(result) => { X(rd) = extend_value(is_unsigned, result); RETIRE_SUCCESS },
    Err(e)     => { handle_mem_exception(vaddr, e); RETIRE_FAIL },
  }
}
\end{lstlisting}
\caption{Sail model for the \CODE{LOAD} RISC-V instruction, together with
some of the functions it needs.}
\label{fig-load}
\end{FIGURE}

The result of calling \CODE{mem\_read} is a \CODE{union} where one of the
variants stores a \CODE{\%bv\_generic} at the JIB level. The width of the result
of the read from simulated main memory is data-dependent on some bits of the
encoded instruction. Therefore it cannot be specialised at compile-time
to the \CODE{width\_bytes} parameter. This also means that the bitwidth of the
result is unknown to the static optimisations.

Similarly, it is much harder to track
concrete bitvector widths when the values pass through a \CODE{union}
or are stored into a \CODE{struct}. It is conceivable to handle such
cases with more powerful intra-procedural static analysis, and taking
the information the liquid types into account. We leave this as future
work, see  Section \ref{sec-future}.

\subsection{RPython code generation}

After Sail functions have been optimised, we generate RPython
source code from each. This is straightforward, the only complexity being
that the IR control flow graphs use basic blocks with (conditional) jumps
between them, and RPython only supports structured control flow. Pydrofoil
compiles that to the standard construction of
a program counter variable and an infinite loop with one condition for
every basic block.\footnote{
    This is an old approach~\cite{bohm_flow_1966},
    see Harel~\cite{harel_folk_1980} for a
    thorough historical discussion. There are much better approaches to
    generating block-structured code out of arbitrary control flow
    graphs~\cite{zakai_emscripten_2011,ramsey_beyond_2022} and we plan to eventually switch to one of
    them.
}


\subsection{Adding a JIT compiler with RPython}
\label{sec_jit}

To gain efficiency beyond what an interpreter-based simulator can offer,
we use RPython's meta-tracing JIT compiler. It traces
the execution of the ISS through one loop of the simulated
guest program and turns the resulting traces into host machine code. In
this way, RPython's tracing JIT acts as a trace-based dynamic binary
translator for the simulated instruction set architecture.

RPython's meta-tracing JIT needs a few annotations in the outermost
execution loop of the Sail model in question. In particular, the
tracing JIT needs to know what the core execution loop is, and which
register of the simulated machine stores the program counter. The
meta-tracing JIT uses the program counter to detect backwards jumps in the guest
program in order to identify loops at the guest program level.

The meta-tracing JIT can further optimise the generic bitvector and integer
operations that the static optimisations of Section~\ref{sec_static_opts} did not
improve. Since it traces a concrete execution path of guest machine
code, all used bitvector widths are observable by the JIT. The JIT can thus
specialise the generic bitvector operations at runtime to the observed
widths. This yields performance improvements because the JIT can remove some of the
remaining heap allocations of bitvectors and integers using escape
analysis~\cite{BolzCF:allrembpeiatj}. However, it is
still preferable to statically remove as many generic bitvector operations as possible,
as otherwise the JIT has to continually remove them, every time the ISS is executed.

Let's look at the JIT trace that is generated for a single RISC-V
\CODE{addi} instruction (the details of this trace are not important):

\begin{minipage}[t]{0.45\textwidth}
\begin{lstlisting}[language=trace]
...
i106 = getfield_gc_i(p1, field=s4)
...
# c.addi s4. 0x8
i144 = getarrayitem_gc_i(ConstPtr(ptr142), 26)
i146 = int_and(i144, 65535)
i148 = int_and(i144, 3)
i150 = int_eq(i148, 3)
guard_false(i150)
i152 = int_eq(i146, 1)
guard_false(i152)
i154 = uint_rshift(i146, 5)
i156 = int_and(i154, 1)
i158 = uint_rshift(i146, 6)
i160 = int_and(i158, 1)
i162 = int_lshift(i156, 1)
i163 = int_or(i162, i160)
i165 = uint_rshift(i146, 7)
i167 = int_and(i165, 15)
i169 = uint_rshift(i146, 11)
i171 = int_and(i169, 3)
i173 = int_lshift(i171, 2)
i174 = int_or(i173, i163)
i176 = int_lshift(i167, 4)
i177 = int_or(i176, i174)
i178 = int_is_zero(i177)
guard_false(i178)
i180 = uint_rshift(i146, 13)
  
\end{lstlisting}
\end{minipage}
\hfill 
\begin{minipage}[t]{0.45\textwidth}
\begin{lstlisting}[language=trace, firstnumber=29]
i181 = int_is_zero(i180)
guard_true(i181)
i182 = int_is_zero(i148)
guard_false(i182)
i184 = int_and(i165, 31)
i186 = uint_rshift(i146, 12)
i188 = uint_rshift(i146, 2)
i190 = int_and(i188, 31)
i192 = int_lshift(i186, 5)
i193 = int_or(i192, i190)
i194 = int_is_zero(i193)
guard_false(i194)
i195 = int_is_zero(i184)
guard_false(i195)
i197 = int_eq(i148, 1)
guard_true(i197)
i199 = int_xor(i193, 32)
i201 = int_sub(i199, 32)
i203 = int_and(i201, 4095)
guard_value(i184, 20)
i206 = int_xor(i203, 2048)
i208 = int_sub(i206, 2048)
i209 = int_add(i106, i208)
i211 = int_add(i2, 2)
i213 = int_add(i0, 2)
setfield_gc(p1, i209, field=s4)
i215 = int_eq(i213, 10000)
guard_false(i215)
# next guest instruction
  
\end{lstlisting}
\end{minipage}

\NI The trace does not contain heap allocations, which is good. However, the trace is
long, given that it performs only a single guest addition. The
problem is the \CODE{getarrayitem\_gc\_i} instruction at the start, which is the
instruction fetch from simulated main memory. A large part of the trace (Lines 6-50) is
manipulating the result of that memory read to perform the instruction
decoding. This happens despite the program counter register being constant for
that point of the trace. In the following section we will look at an optimisation for
removing those \emph{instruction-decoding} trace instructions.

\subsection{Making the main memory simulation JIT-friendly}
\label{sec_mem_immutable}

One big problem for the JIT compiler is that while it knows the
concrete value of the guest program counter, that is not
enough for optimising away the decoding of the instruction stored at
that program counter.

The reason is the possibility of \EMPH{self-modifying code}.
In practice true self-modifying code is rare, and so we want the JIT to exploit this fact.
To do this, we track the status of every part
of the simulated main memory.  Currently, this tracking happens at the granularity of 8
bytes.\footnote{It is possible that other granularities work better.
Pydrofoil could \EG do this kind of tracking on the granularity of guest pages
and take the permission bits from the page tables into account.}
Every 64-bit word in simulated main memory can be in one of three
states: \CODE{status\_default}, \CODE{status\_immutable},
\CODE{status\_mutable}. All the 64-bit words start out in
\CODE{status\_default} state. If a \CODE{status\_default} 64-bit word is read
or written, its status is unchanged, and the memory read proceeds normally. However, if a
\CODE{status\_default} is read from during instruction fetch, it
transitions to \CODE{status\_immutable}. The JIT will constant-fold
reads from \CODE{status\_immutable} 64-bit words and return the concrete
value of the bits stored in the 64-bit word.
To behave correctly in the face of self-modifying code,
the simulated main memory checks on every write whether a
\CODE{status\_immutable} 64-bit word is being written to.\footnote{This
extra check makes all writes to simulated main memory slower, but the
trade-off is worth it because instruction fetch is common:
there are more instruction fetches than memory writes.}
When a
\CODE{status\_immutable} 64-bit word
is changed, the generated host machine code is potentially no longer
valid, because it was generated under the assumption that
\CODE{status\_immutable} 64-bit words are immutable. Therefore all relevant
host machine code is invalidated in such a situation, and the modified
64-bit word is marked as \CODE{status\_mutable}. \CODE{status\_mutable} 64-bit
words are assumed to potentially contain mutable code, so the JIT will never
constant-fold reads from them during host machine code generation.
Figure \ref{image_2025-pydrofoil-mem-states} shows a state diagram of
the various states every 64-bit word of the simulated main memory can be in, and the
transitions it can go through.

\begin{FIGURE}
  \IMAGESIZE{0.5}{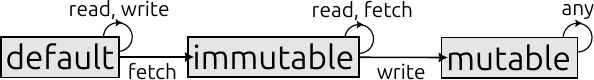}
  \caption{Diagram of the states a 64-bit word of simulated main
    memory can be in.}
  \label{image_2025-pydrofoil-mem-states}
\end{FIGURE}

After adding this status tracking logic to the simulated main memory, the trace
for the \CODE{addi} instruction in Section~\ref{sec_jit} gets much simpler.
Since the memory read of the instruction bits happens from a
\CODE{status\_immutable} 64-bit word, the bits of the guest instruction are
constant. Therefore the trace operations that decode those bits all get
constant-folded away. This kind of trace length reduction due to memory status
tracking is typical: on average the trace for one guest instruction gets 3--4
times shorter with memory status tracking. The remaining trace looks like this:

\begin{lstlisting}[language=trace]
i35 = getfield_gc_i(p1, field=s4)
...
# c.addi s4. 0x8
i44 = int_add(i35, 8)
i64 = int_add(i0, 2)
setfield_gc(p1, i44, field=s4)
i66 = int_eq(i64, 10000)
guard_false(i66)
\end{lstlisting}

\NI Now the trace looks good! The \CODE{addi\ s4, 0x8} instruction
is compiled to a \CODE{int\_add(..., 8)} instruction, which is its
closest equivalent in the RPython trace IR. There are a few extra
instructions around to read and write the value of the register field
in the \CODE{p1} \CODE{struct} which contains the simulated machine's state. In
addition, there are three instructions that check whether 10,000
instructions have been executed, which would make it time to tick the
simulated clock (we discuss the issue of clocks in more details in Section
\ref{sec-future}).

\subsection{Improving the RPython JIT integer optimisations}
\label{sec-rpython-jit-int-opts}

While developing Pydrofoil, we identified
a number of missing trace optimisations for bit manipulation
operations in the RPython JIT.  This may not be surprising: the RPython JIT was developed
for dynamic language implementations where such operations are rare.
In contrast, bit manipulation is pervasive in processors, and, as a
consequence, in the Pydrofoil traces the JIT needs to optimise.  In order to
adapt the JIT's optimising passes to such workloads, we have added a
number of new integer optimisations. Let us give an example.  In
processors, individual bits are often used to hold permissions, \EG
whether a specific address range allows non-aligned access. Typically
such a permission bit rarely changes, if ever. Nevertheless the
permission bit is checked on every access. In hardware such a check is
cheap.  In software simulation of hardware, checking the bit again and
again is expensive. In principle, a JIT can determine during
trace-optimisation if a permission bit changes from one instruction to the next. If not, all
but the first check can be elided. However JITs targeting high-level languages
typically do not track value-constancy \EMPH{at the bit-level}.

We adapt
\EMPH{Tristate numbers}, used in the Linux kernel by the eBPF verifier to
reason about which bits in a variable have constant values~\cite{vishwanathan_tristate_2022},
to the RPython JIT. An equivalent approach
is also used in other compilers such as LLVM, where it is called
\EMPH{Known Bits}~\cite{taneja_knownbits_2020}.  The
Tristate number abstraction lets the JIT track values of constant bits of interest
in integer and bitvector variables (which map to the same JIT trace IR type)
when optimising traces.
This enables the removal  of  permission checks where the JIT can prove that the
relevant bits cannot change.

A concrete example for why this is useful are alignment checks. If two load
instructions are executed that use the same base register, e.g.\ like this:

\begin{lstlisting}[language=CCustom]
ld t0, 0x8(a0)
ld t1, 0x10(a0)
\end{lstlisting}

Then both loads have to check whether the pointer the value comes from is word-aligned, which
means that the lowest bits are zero. The Tristate number analysis lets the JIT
conclude that if the first load is aligned, the second one is aligned too, and
so the alignment check can be removed from the second load. Here is a
sketch\footnote{The actual trace is significantly more complicated, due to
bounds checks and address translation.}
of what this looks like in a JIT trace:

\begin{lstlisting}[language=trace]
# ld t0, 0x8(a0) # first load
i1 = getfield_gc_i(p1, field=a0)
i2 = int_add(i1, 8) # add offset to get the actual pointer
i3 = int_and(i2, 7) # alignment check: extract last three bits
i4 = int_is_zero(i3) # check that they are zero
guard_true(i4) # after this instruction the JIT knows that i1 and i2 are of the form ???...???000
... # do the actual load from memory

# ld t1, 0x10(a0) # second load
i12 = int_add(i1, 16) # must be of the form ???...???000, because both i1 and 16 are
i13 = int_and(i12, 7) # alignment check for second load: i13 is 0
i14 = int_is_zero(i13) # always 1
guard_true(i14) # can be removed by the JIT's opimizer because i14 is constant 1
... # do the actual load from memory


\end{lstlisting}

\section{Benchmarks}\label{section_benchmarks}

To evaluate the performance of Pydrofoil and quantify the impact of
our domain-specific optimisations and JIT compilation, we conducted a
comprehensive benchmarking study.  Our evaluation, constrained by
available resources, aims to be a representative set of industrial ISS
workloads, and has four main parts.

\begin{itemize}

\item Comparison of Pydrofoil with the Sail-generated ISS, using RISC-V 64.

\item Ablations to understand the contribution of each individual
  domain-specific Pydrofoil optimisation, using RISC-V 64.

\item Comparison of Pydrofoil with QEMU, a handwritten
  industry-standard ISS.

\item Preliminary benchmarks using Arm v9.4a and CHERIoT ISAs.

\end{itemize}

\NI We use two classes of benchmark: SPEC CPU and booting Linux (with
the exception of CHERIoT, described below).  Both are widely used in
computer architecture, and motivated next.
\begin{itemize}

\item SPEC, the Standard Performance Evaluation Corporation benchmark
  \cite{SPEChomepage}, is a suite of standardised tests evaluating the
  performance of computer systems.  We focus on the SPEC CPU
  benchmarks \cite{BucekJ:specpungcb}, which measure the performance
  of a system's processor and memory subsystem.
  We restrict our attention to SPECint, and avoid the floating point
  arithmetic in SPEC CPU because currently the RISC-V model lacks a
  Sail implementation of the floating point extensions. Instead, the
  RISC-V model relies on the C-based Berkeley SoftFloat library
  \cite{HauserJ:softfloats}, and benchmarking floating points would
  \EMPH{not} exercise Pydrofoil\footnote{A pure Sail implementation
  of the floating point extensions is in preparation. When it becomes
  part of the official RISC-V model, it makes sense to revisit our
  benchmarks.}.  See Appendix \ref{app_benchmarks} for more detail
  about the individual SPEC benchmarks used.

\item Booting Linux is a critical benchmark due to its prevalence in
  ISS applications and its challenge for JIT compilers, as the boot
  process lacks the long-running loops that JITs excel at optimising.
  Typically, the boot process executes a lot of code that is never
  seen again, and so this benchmark helps to establish JIT compilation
  latency.

\end{itemize}

\NI Each SPEC benchmark was integrated into a complete Linux 6.7.0
boot image, using the initial ramdisk~\cite{enwiki:1277729657} (\texttt{initrd}). This
approach is necessary because the Sail models do not contain emulated
disk devices that could have been used used for storing
programs. Unfortunately, due to the limitation of a few hundred MB on \CODE{initrd}
sizes, we were not able to run \CODE{657.xz\_s.1}. For the \CODE{625.x264}
variants we had the same problem but moved input file generation into the Linux
image.  We mark those SPEC benchmarks with an asterisk (*) when we discuss them in
Section~\ref{sec_bench_comparison}.

Due to the deterministic
nature of Sail models, and their lack of access to the host clock, all timing
was performed externally on the host system, measuring the simulator's
total wall-clock runtime. Therefore, our SPEC benchmark times also include the Linux 6.7.0
boot process until user-mode (about 138 million instructions).

\PARAGRAPH{Hardware and software used for benchmarking}
All Sail compilation uses the following parameters affecting
optimisation:
\CODE{-\!dno\_cast\ -\!O\ -\!Oconstant\_fold\ -\!memo\_z3}.  Generated
C was compiled with \CODE{gcc}.  For pragmatic reasons, we use three
different compute setups (indicating for each benchmark which is
used).
\begin{itemize}

\item TaiShan server with Kunpeng 920 Arm v8.4 cores, 500 GB main memory, running Ubuntu 18.04.6.

\item AMD Ryzen 7 PRO 7840U with 32 GB main memory, running Ubuntu 24.04.6.

\item AMD Ryzen 9 7950x3d, with 128GB main memory, running Ubuntu  24.04.1.

\end{itemize}

\subsection{Comparing Pydrofoil and Sail performance for the RISC-V model}

Pydrofoil's most important goal was generating faster ISSs than Sail
itself. This section compares Pydrofoil with Sail. We are
resource constrained: the Sail-generated RISC-V ISS is too slow!  For
example \CODE{605.mcf\_s.0}, which executes 1.2 trillion instructions, takes
roughly six months on the TaiShan server.
Therefore, we decided to run the comparison on only two arbitrarily
chosen SPEC benchmarks, \CODE{605.mcf.0} and \CODE{600.perlbench.0}.
We stopped the benchmarks for the Sail-generated ISS after 20 billion
instructions. Pydrofoil is run to conclusion on the
same benchmarks.  We compare the instructions per second executed by
both systems, extrapolated from the first 20 billion for Sail.

\begin{FIGURE}
 \IMAGE{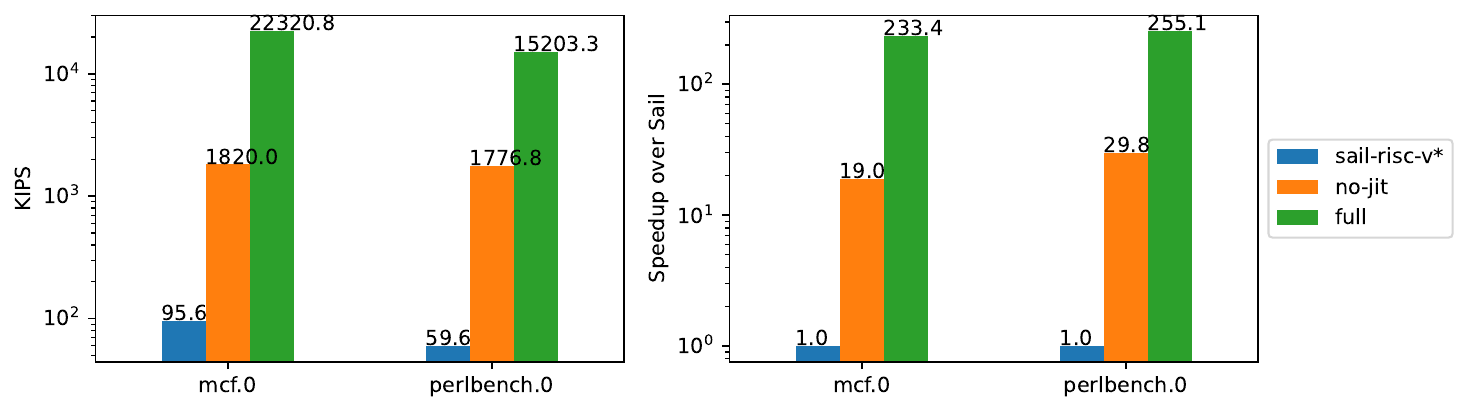}
  \caption{Instructions/second comparison between Sail and Pydrofoil. The left
    plot shows KIPS (= kilo instructions per second),
    the right the speedup over Sail. Higher is better.}
  \label{image_sail_comparison}
\end{FIGURE}

\PARAGRAPH{Results}
This benchmark uses the TaiShan server.  The results of this can be
seen in Figure~\ref{image_sail_comparison}. We see that Pydrofoil is
233$\times$ and 255$\times$ faster than the Sail RISC-V ISS on both
benchmarks. Even without the JIT, Pydrofoil is at least 19$\times$ faster
than Sail.

\subsection{Ablations: benchmarking Pydrofoil variants}\label{bench_ablations}

To quantify the effect of our optimisations, we conduct ablation
studies: we systematically disable successively more optimisations and
measure the resulting slowdowns. This gives us a quantitative
approximation of the relative impact of each optimisation, providing
insights into their individual and collective value. We use the
following variants.
\begin{itemize}
 
\item \CODE{full}: The full Pydrofoil simulator, running the JIT and
  the AOT with all optimisation enabled.

\item \CODE{no\!-\!mem\!-\!immutability}: Disables memory immutability
  tracking (Section \ref{sec_mem_immutable}).

\item \CODE{no\!-\!jit}: Additionally disables the JIT (Section
  \ref{sec_jit}) and works only as an interpreter at runtime.

\item \CODE{no\!-\!bv\!-\!int\!-\!opts}: Additionally disables the
  static optimisations that replace generic bitvector and generic
  integer operations with more specialise variants operating on 64-bit
  integers (Section \ref{sec_static_opts_bv_int}).

\item \CODE{no\!-\!spec}: Additionally disables inlining (Section
  \ref{sec_inlining}) and function specialisation (Section
  \ref{sec_specialisation}).

\item \CODE{no\!-\!static\!-\!opt}: Additionally disables all
  remaining static optimisations (Section \ref{sec_static_opts}).

\end{itemize}

\NI Unfortunately, we were unable to run all ablations against the
SPEC CPU benchmarks listed in Appendix \ref{app_benchmarks} as
planned.  The problem is that the ablations with many of the
optimisations disabled run much to slow to finish in a realistic time
frame.  For example the aforementioned \CODE{605.mcf\_s.0}
takes several months with \CODE{no\!-\!static\!-\!opt}.
We deal with this as we did in the previous comparison and run the
ablations on the same arbitrarily chosen SPEC benchmarks,
\CODE{605.mcf.0} and \CODE{600.perlbench.0}.  We stopped the
benchmarks for the slowest ablations after 20 billion
instructions. The remaining ablations are run to conclusion on the
same benchmarks.  We compare the instructions per second executed by
both systems, extrapolated from the first 20 billion for the slowest
ablations.

\begin{FIGURE}
  \IMAGE{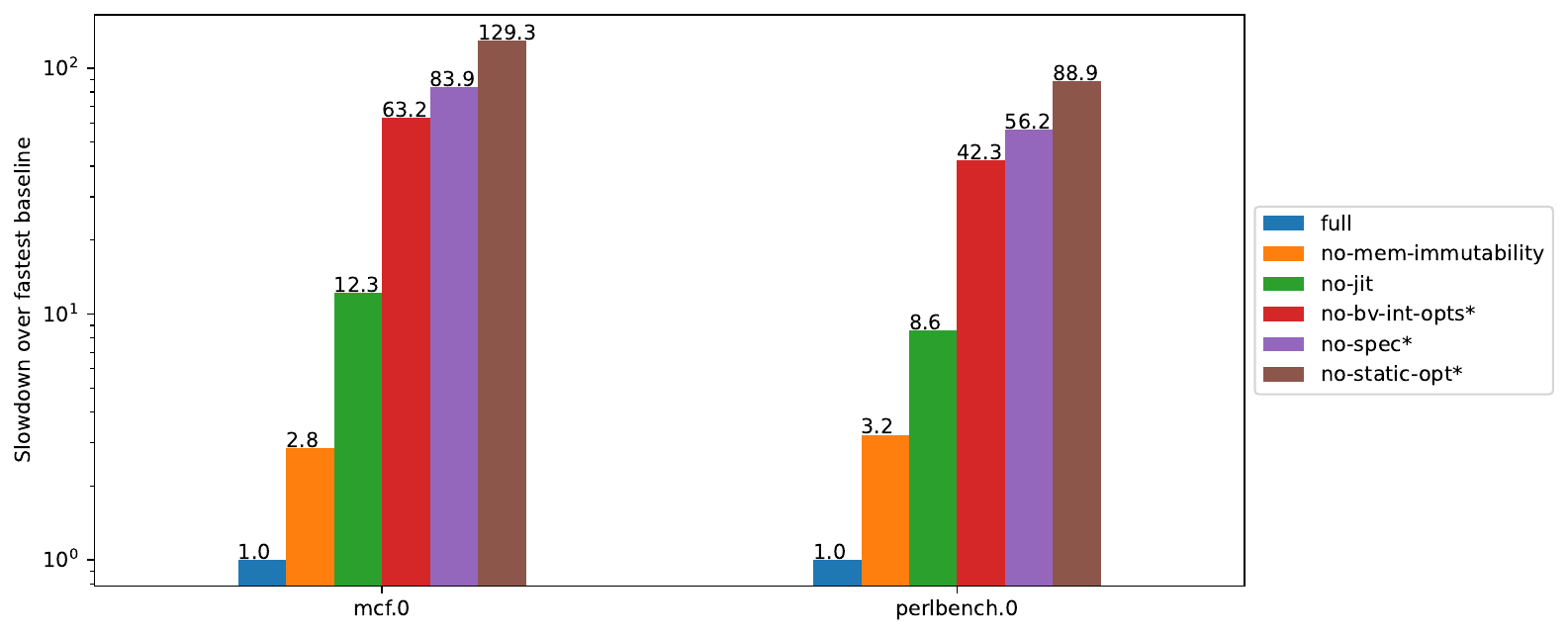}
  \vspace{-1.5em}
  {\small
\begin{tabular}{lrrrr}
\toprule
 & \multicolumn{2}{c}{KIPS} & \multicolumn{2}{c}{slowdown} \\
Benchmark & mcf.0 & perlbench.0 & mcf.0 & perlbench.0 \\
Implementation &  &  &  &  \\
\midrule
full & 22320.79 & 15203.28 & 1.00 & 1.00 \\
no-mem-immutability & 7836.32 & 4701.09 & 2.85 & 3.23 \\
no-jit & 1820.04 & 1776.76 & 12.26 & 8.56 \\
no-bv-int-opts* & 353.13 & 359.32 & 63.21 & 42.31 \\
no-spec* & 266.06 & 270.47 & 83.89 & 56.21 \\
no-static-opt* & 172.60 & 170.92 & 129.32 & 88.95 \\
\bottomrule
\end{tabular}
}

\caption{Benchmark results for ablations, as graph and table. The y-axis of the
    plot is logarithmic. Benchmark results with * are extrapolated from the
    first 20 billion instructions.}
  \label{fig-ablations}
\end{FIGURE}

\PARAGRAPH{Results}
This benchmark uses the TaiShan server.  The results are shown in
Figure \ref{fig-ablations}.  We see
clearly that the JIT is fundamental for performance, it provides
(together with memory immutability tracking) a 12$\times$/8$\times$ performance
boost. The static bitvector and integer optimisations are the most
useful static optimisations. Disabling them slows Pydrofoil down by about a further
5$\times$. Inlining, function specialisation and all other static
optimisations are less important. Turning the runtime-switchable
integer representation off leads to a further slowdown of more than
2$\times$.

We also investigated what is more important: static optimisations or
JIT compilation.  To that end, we add another ablation,
\CODE{no\!-\!opt\!-\!with\!-\!jit} It is outside of the sequence of
ablations in the previous subsection, because it disables all static
optimisations but leaves the JIT-compiler active. The results are in
Figure~\ref{image_jit_ablations}. We see that this variant is
4.54$\times$/5.25$\times$ slower than \CODE{full}. This slowdown is less
than \CODE{no\!-\!jit}, which removes the JIT but keeps the static optimisations.
We conclude that the JIT can partly
compensate for missing static optimisations (but not vice versa). The
JIT does this by performing some, but not all, of the missing static
optimisations at runtime.

\begin{FIGURE}
  \IMAGE{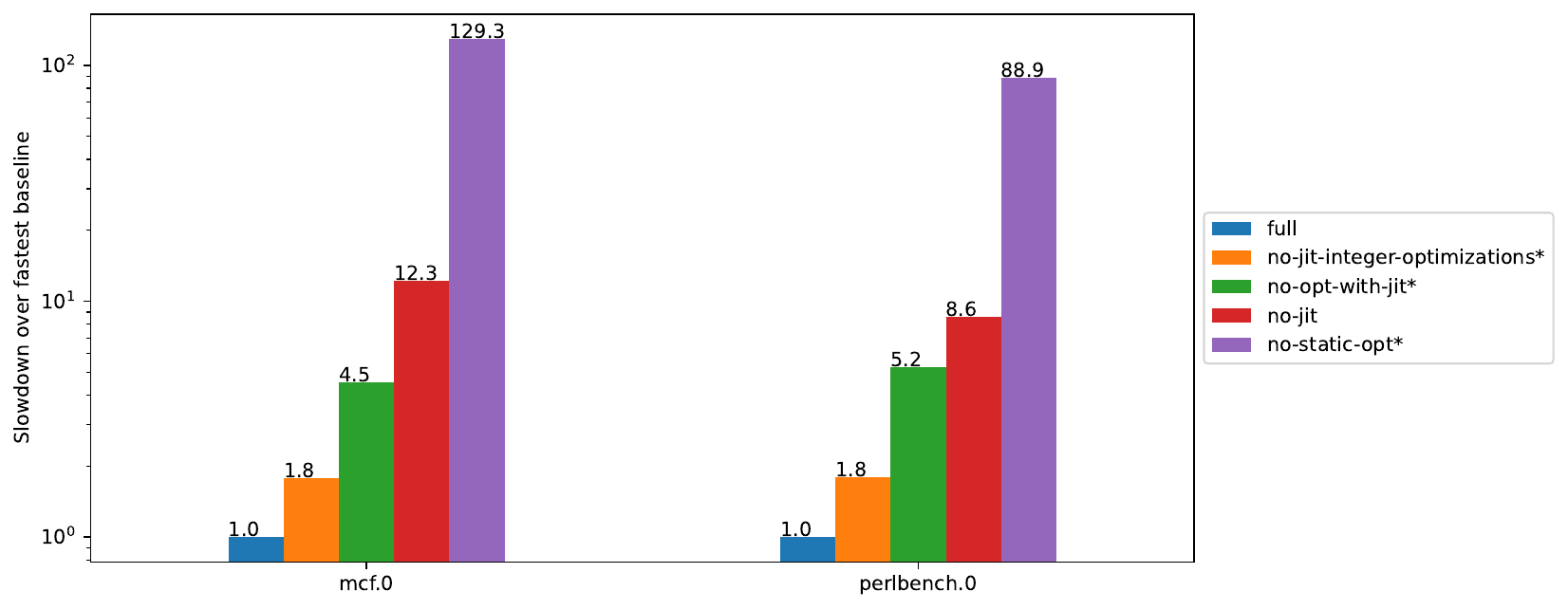}
  {\small
\begin{tabular}{lrrrr}
\toprule
 & \multicolumn{2}{c}{KIPS} & \multicolumn{2}{c}{slowdown} \\
Benchmark & mcf.0 & perlbench.0 & mcf.0 & perlbench.0 \\
Implementation &  &  &  &  \\
\midrule
full & 22320.79 & 15203.28 & 1.00 & 1.00 \\
no-jit-integer-optimisations* & 12488.84 & 8454.55 & 1.79 & 1.80 \\
no-opt-with-jit* & 4912.80 & 2898.55 & 4.54 & 5.25 \\
no-jit & 1820.04 & 1776.76 & 12.26 & 8.56 \\
no-static-opt* & 172.60 & 170.92 & 129.32 & 88.95 \\
\bottomrule
\end{tabular}
}

  \caption{JIT-relevant ablation results. The y-axis of the
    plot is logarithmic.}
  \label{image_jit_ablations}
\end{FIGURE}

Unfortunately we were unable to benchmark the additional integer
optimisations from Section~\ref{sec-rpython-jit-int-opts}: Tristate
numbers are now too embedded into the RPython JIT to be easily disabled.
To get an upper bound of the effect those optimisations have, we ran
an ablation \CODE{no\!-\!jit\!-\!integer\!-\!optimisations} that
disables \EMPH{all} integer optimisations of the RPython JIT (also in
Figure~\ref{image_jit_ablations}). Doing
that gives  a slowdown of 1.79$\times$ and 1.80$\times$. Since the RPython JIT also
had integer optimisations before the changes motivated by Pydrofoil
(including a range analysis done by the JIT at runtime), we expect the
effect of Tristate numbers to be below that number. Therefore it is a
less important optimisation compared to any of the Pydrofoil static
optimisation and compared to having a JIT in the first place.

\begin{FIGURETOP}
  \begin{tabular}{lrrr}
\toprule
Benchmark & time pydrofoil & time qemu & speedup \\
\midrule
600.perlbench\_s.0 & 25462s & 1121s & 22.7$\times$ \\
600.perlbench\_s.1 & 49089s & 760s & 64.6$\times$ \\
600.perlbench\_s.2 & 14026s & 692s & 20.3$\times$ \\
602.gcc\_s.0 & 31757s & 993s & 32.0$\times$ \\
602.gcc\_s.1 & 35479s & 651s & 54.5$\times$ \\
602.gcc\_s.2 & 30758s & 627s & 49.1$\times$ \\
605.mcf\_s.0 & 13248s & 6095s & 2.2$\times$ \\
620.omnetpp\_s.0 & 58203s & 1267s & 45.9$\times$ \\
623.xalancbmk\_s.0 & 14044s & 3456s & 4.1$\times$ \\
625.x264\_s.0* & 15664s & 278s & 56.3$\times$ \\
625.x264\_s.1* & 92698s & 947s & 97.9$\times$ \\
625.x264\_s.2* & 69464s & 768s & 90.5$\times$ \\
631.deepsjeng\_s.0 & 39604s & 1749s & 22.6$\times$ \\
641.leela\_s.0 & 34928s & 1738s & 20.1$\times$ \\
648.exchange2\_s.0 & 17023s & 1159s & 14.7$\times$ \\
657.xz\_s.0 & 49566s & 2603s & 19.0$\times$ \\
Geometric mean &  &  & 26.7$\times$ \\
\bottomrule
\end{tabular}

    \caption{Performance results of SPEC benchmarks on QEMU and Pydrofoil, in
    seconds. Speedup is the factor by which QEMU is faster than Pydrofoil. All
    benchmark times include booting Linux. Benchmarks marked with * also include
    the time to generate their input files.}\label{fig-spec-qemu}
\end{FIGURETOP}

\subsection{Comparison with QEMU}
\label{sec_bench_comparison}

To assess Pydrofoil's real-world usability, we benchmark its
performance against an industrial-strength ISS.  Given the limited availability
of ISS generators that can compile Sail, that means using one not
based on Sail. We choose QEMU~\cite{BellardF:qemfasapdt}, a widely
used, manually optimised C implementation. QEMU supports numerous
prominent ISAs, including RISC-V, and serves as a component in various
processor simulation frameworks.  QEMU's core (JIT) compilation
engine translates guest instructions
into host instruction sequences with a hand-crafted, ISA-specific
translator.  Over two decades of extensive engineering has contributed
to QEMU's exceptional simulation performance. While not designed for
formal verification, its widespread adoption positions QEMU as a
practical performance ceiling for Pydrofoil.

We use SPEC and booting Linux as benchmarks, and RISC-V as ISA.

\PARAGRAPH{Results}
This benchmark uses the AMD Ryzen 9 setup. The results are in Figure~\ref{fig-spec-qemu}.
We see that QEMU's performance is significantly ahead of
Pydrofoil's: QEMU is between 2$\times$ and 100$\times$ faster than Pydrofoil, with a
geometric mean of \qemuspeedup. This is clearly disappointing, it
shows that for many cases Pydrofoil is not good enough at removing the
overhead of Sail. Section~\ref{sec-future} gives concrete suggestions
on how to reduce the performance gap with QEMU.

\subsection{Preliminary benchmarks for Pydrofoil-Arm and Pydrofoil-CHERIoT}

All benchmarks above use RISC-V. This raises the question how
dependent on RISC-V the performance numbers are.  This benchmark
presents preliminary performance results for two other ISAs, Arm
v9.4-a~\cite{SailArmRepo} and CHERIoT~\cite{CHERIoTPlatform,SaarA:checommsfed,CHERIoT:arcspec}.
For both
ISAs, we compared the performance of the Pydrofoil-generated ISS with
that of the Sail C backend.

The Arm v9.4-a specification in Sail is obtained by transpilation from
Arm's ASL~\cite{ReidA:truspeoaveaavamsla} specification, and is
\EMPH{substantially} bigger and more complex than RISC-V. For example
the JIB obtained by translation of the Arm model is 20$\times$
larger than the corresponding RISC-V.  As Arm v9.4-A benchmark, we
boot Linux 6.0.7, but only up to the initialisation of the \CODE{init}
process. This executes 24.6 million instructions.

In contrast to the complex Arm ISA, CHERIoT is a simple 32-bit RISC-V
micro-controller without virtual memory and caches, co-designed with a
real-time operating system (RTOS), all with a strict focus on security.
CHERIoT is a capability machine, based on (a small variant of) CHERI~\cite{watson:cheriisav9}.
Capabilities can be seen as a hardware
implementation of 'fat pointers': conventional integer pointers are
replaced by capabilities, which are memory addresses together with
permissions constraining capability use.  This enables the processor
to check every memory access to ensure that it is not out-of-bounds
(spatial memory safety).  Capabilities can be used to mitigate a wide
range of memory safety bugs and vulnerabilities.  CHERIoT comes with a
Sail specification of its ISA semantics~\cite{CHERIoT:speandsm}, and
the ISS that Sail generates is currently the only available ISS for
CHERIoT.  For CHERIoT we boot its RTOS~\cite{CHERIoTPlatform:chrrto}
and then run CHERIoT's ``allocation-benchmark''~\cite{CHERIoTPlatform:alloc_bench}, see Appendix
\ref{app_benchmarks} for details.
Booting the RTOS takes 490,000 instructions, while the ``allocation-benchmark''
takes 274 million.

\PARAGRAPH{Results}
This benchmark uses the AMD Ryzen 7 setup.  The results for Arm are in
Figure~\ref{figure_measurements_arm}. While Pydrofoil still manages to
produce a much faster ISS for Arm than the Sail C backend (by a factor
of almost 70$\times$), in terms of instructions per second the
performance is significantly worse than that of the RISC-V variant.

The CHERIoT results are in Figure~\ref{figure_measurements_cheriot}.
The speedup of Pydrofoil-CHERIoT is
about 50$\times$, even lower than that of Arm. This seems to be at
least partly caused by Pydrofoil's poor handling of the fact that the
CHERIoT-Sail-model uses a Sail \CODE{struct} type to store the CHERIoT
capability registers (in RISC-V and Arm most registers are
bitvectors). We hope to improve this problem in the future.

\begin{FIGURE}
    \IMAGE{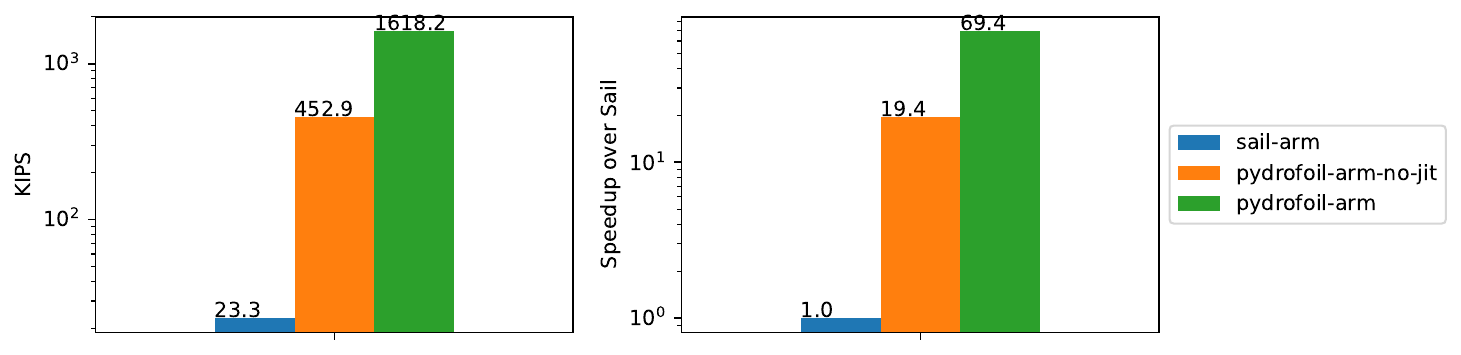}
  \caption{Benchmark results for the Arm ISS, booting Linux. The left plot
    shows KIPS, the right the speedup over Sail. Higher is
    better.}\label{figure_measurements_arm}
\end{FIGURE}

\begin{FIGURE}
    \IMAGE{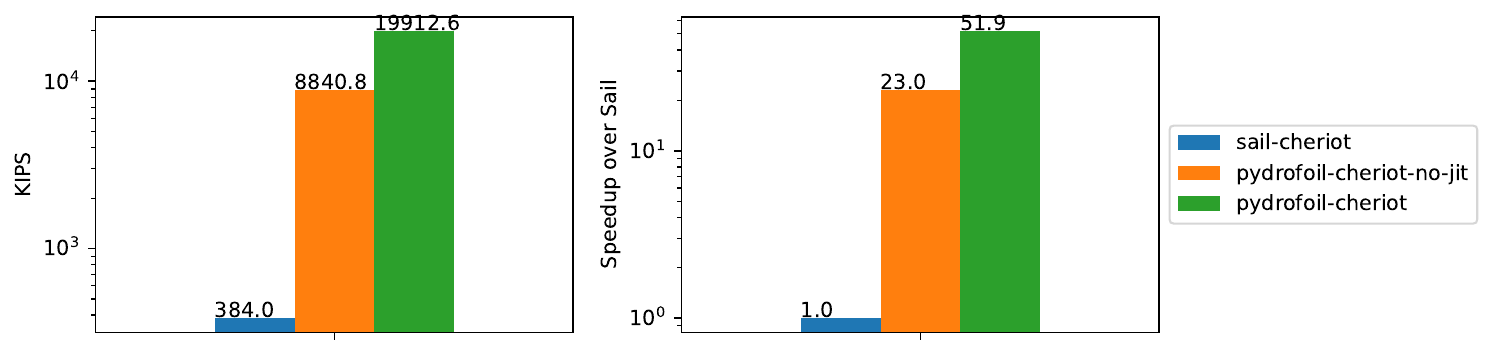}
    \caption{Benchmark results for the CHERIoT ISS, running the
    allocation-benchmark. The left plot shows KIPS, the right the speedup over
    Sail. Higher is better.}\label{figure_measurements_cheriot}
\end{FIGURE}

\subsection{Threats to validity}\label{section_bench_threats}

This study, while providing insights into Pydrofoil's performance, is
subject to several threats to validity. We acknowledge these
limitations and discuss them in detail next.

The primary limitation stems from the computational intensity of
benchmarking ISSs, which operate at significantly lower speeds than
the hardware they model. A comprehensive evaluation of Pydrofoil
across a broad range of benchmarks requires computational resources
and time that exceed our current capacity. Consequently, our findings
are based on a subset of representative benchmarks, potentially
limiting the scope of our conclusions.  Second, the selected
benchmarks, while intended to be representative of typical computer
architecture workloads, were not specifically designed for ISS
evaluation. This misalignment may not fully exercise Pydrofoil's
capabilities, potentially under-representing its strengths and
weaknesses.  We acknowledge the need for benchmarks specifically
tailored to evaluate ISS generators.  Third, to mitigate limitations
in existing tools, we employed certain experimental
shortcuts. Notably, we did not perform multiple benchmark runs to
assess statistical variation\footnote{Due to the required computational intensity.
We hope to fix this in later versions of this work.}.
Finally, our primary focus on the RISC-V ISA introduces a
potential limitation to the generalisability of our findings. While we
included preliminary evaluations for Arm and CHERIoT, these are
subject to methodological constraints.  The Arm Sail model, derived
from ASL through transpilation, may introduce runtime checks due to
ASL's weaker typing, potentially confounding the attribution of
performance characteristics. This highlights the challenges of
comparing ISSs derived from different specification languages.  The
CHERIoT evaluation is based on a single, small-scale benchmark, whose
results may not generalise well.

We nevertheless believe that our work nevertheless demonstrates that
ISA specifications in Sail can be compiled into performant ISSs.

\subsection{Summary of benchmarking}

To address research questions RQ2 and RQ3, we conducted extensive
benchmarking, aiming to quantify Pydrofoil's performance. Our ablation
studies demonstrate that Pydrofoil's hybrid architecture, combining
AOT and JIT compilation with domain-specific optimisations, is a
viable approach for accelerating ISSs generated from Sail.  Our
smaller-scale benchmarks across Arm and CHERIoT ISAs corroborate the
trends observed with RISC-V, giving us additional confidence in the
robustness of our approach.  However, comparisons with QEMU highlight
a significant remaining performance gap, indicating substantial room for
improvement to match the speed of hand-crafted full-system emulators.

Notably, the considerable resource demands of these benchmarks suggest
a need for more efficient and feasible benchmarking strategies.

\section{Suggestions for future research}\label{sec-future}

While the present work has demonstrated significant performance gains,
achieving a >250$\times$ speedup over the Sail-generated ISS, Pydrofoil
remains \qemuspeedup slower than QEMU.
Spink et al.\ (2020)~\cite{SpinkT:retsysldbth} report a 2$\times$ speedup over QEMU
by employing a combination of AOT and partial evaluation techniques,
similar in principle to Pydrofoil, albeit compiling from a low-level
ADL, which is easier to optimise for than Sail.  These discrepancies
highlight the potential for further optimisation, and motivate the
roadmap presented below, which is intended to bring Pydrofoil to
performance parity with QEMU.  It might also provide insights for the
development of high-performance ISS generators for other high-level
ADLs.

\PARAGRAPH{Clocks}
Simulation of the processor clock is a known inefficiency for ISS
acceleration~\cite{SpinkT:effasyihiafsiss}. The RISC-V Sail model
ticks its clock every 100 instructions. This number was chosen
arbitrarily, and is \EMPH{not} part of the RISC-V instruction set
architecture~\cite{WatermanA:riscvISAvolI,WatermanA:riscvISAvolII}.
Checking whether the next clock tick is reached in every guest
instruction is expensive. Moreover, the JIT must generate control flow
edges from \EMPH{all} instruction to the clock logic.  QEMU optimises
this by only checking for clock ticks at the start of generated basic
blocks, and deoptimises whenever a timer event happens elsewhere
\cite{QEMU:tcginsc}.  It is interesting to investigate if QEMU's
approach or the mitigations discussed in~\cite{SpinkT:effasyihiafsiss}
improve Pydrofoil performance.

\PARAGRAPH{Address translation} Real processors cache the results
of address translation in the translation lookaside buffer (TLB), a
dedicated hardware cache.  Simulation has no TLB, so every memory
access must simulate a full page table walk.  This is a known problem
in ISS research~\cite{BellardF:qemfasapdt,SpinkT:haracccafsv}.  We
believe that a JIT with suitable domain-specific optimisations can
lower the cost of simulated page table walks by exploiting their
predictability.

\PARAGRAPH{JIT warmup} The time it takes the JIT to generate optimised
machine code is called warmup~\cite{BarrettE:virmacwbhac},
and is the result of a fine balance between the cost and benefits of
optimisations.  PyPy's tracing mechanism was optimised for general
purposed workloads: in PyPy, one Python bytecode produces 6-7 trace
operations on average, which the optimiser reduces to about 2.
Contrast this with Pydrofoil: tracing one RISC-V guest instruction produces
more than 600 trace operations, which are optimised down to about 65
on average.  The trade-offs driving the JIT optimisation design in
PyPy are likely to be suboptimal for ISS workloads, hence have
significant potential for improvement.

\PARAGRAPH{Type-erasure}
Sail's liquid types can express strong constraints on 
bitvector widths. Sail's current compilation chain mostly erases them.
Consider \CODE{size\_bytes} from Section~\ref{sec_implementation_when_fail}
with signature \CODE{size\_bytes : word\_width \leftrightarrow \{1, 2, 4, 8\}}. In other words, any
value obtained by calling \CODE{size\_bytes} is constrained to values $1
, 2 , 4$ or $8$ in the Sail source.  This strong constraint is lost in
JIB: all the Pydrofoil AOT sees is that \CODE{size\_bytes} returns
\CODE{\%i64}.  If we want to optimise further, the fact that \CODE{\%i64}
stores one of $1, 2, 4,
8$ needs to be re-discovered by static analysis, either at compile-time,
or by the JIT at run-time.  Such strong constraints are prevalent in ISAs.
Given the increasing interest in liquid types in other languages~\cite{VazouN:liqhashaatp,LehmannN:fluliqtfr},
we believe that a better exploitation of the information from liquid types in the compilation toolchain might
deliver better optimisations even beyond Sail.


\section{Related work}

\subsection{DSLs used in computer architecture}
There are a variety of DSLs that have been created for processor descriptions.
Most relevant here are the aforementioned \EMPH{architecture description languages}.
Examples include nML~\cite{FaithA:desinsspun}, ISDL~\cite{HadjiyiannisG:isdlinssdlfr},
EXPRESSION~\cite{HalambiA:explanfaetcsr}, SystemC~\cite{IEEE:ieestafsslrm},
ArchC~\cite{RigoS:arcsyscbadl}, and GenC~\cite{WagstaffH:frohigladtfiss}.  They are
often described as being high-level, and are actively used to derive ISSs, especially in
commercial tools.  Compared to Sail, and with the exception of GenC, they all betray
their origins in layers of the processor design stack lower than ISA: they invariably
also contain facilities to describe low-level detail, whether pipeline
layout, timing constraints, or data-path conflicts.  In their context
of origin, computer architecture, they are considered higher-level,
but only by contrast with RTL languages, such as Verilog and VHDL.  Such DSLs are often
based on C/C++, a notable exception being SSL~\cite{CifuentesC:spesemomi},
which specifies the ISA semantics of machine instructions in Object-Z (an
extension of Z specification language).  However, this approach has
not gotten much traction.  It is only with Arm's ASL~\cite{ReidA:truspeoaveaavamsla},
Sail's most important predecessor, that a mature, ISA-only DSL emerged.

\subsection{ISS acceleration}
Evaluating Pydrofoil's performance against existing ISS solutions
presents challenges due to the lack of directly comparable systems
that utilise Sail.
Proprietary tools such as Arm's FastModel~\cite{Arm:fasmoddswh} and
Architecture Envelope Models~\cite{Arm:armarchfvps}, offer
high-performance simulation capabilities. However, their internal
architectures, including the extent to which they are automatically
generated from Arm's ASL, remain largely undisclosed. Similarly,
commercial tools from Synopsys (\EG ASIP Designer~\cite{Synopsys:asides},
Platform Architect~\cite{Synopsys:plaarcsaaaofpap},
ImperasDV~\cite{ImperasSoftware:riscomffcrvs,Synopsys:imprisvpvme}),
MachineWare~\cite{BosbachN:towhigpvpapsfstcm,MachineWare:vircommlv}, and
Codasip~\cite{Codasip:whatiscodal} rely on handwritten components or
generation from lower-level ADLs like SystemC, nML, or Codal \cite{Codasip:whatiscodal}.
Many academic ISSs, while valuable, have not seen recent
publications or updates to their codebase, making comparison difficult.


\PARAGRAPH{Pydgin}
Pydgin~\cite{LockhartD:pydgenfissfsadwmtjc}, another meta-tracing and
PyPy-based ISS, simulates only user-mode execution.  This
simplification omits address translation, self-modifying code, and
interrupt handling, and utilises the Pydgin Architecture Description
Language, which lacks the abstraction level of Sail. These differences
render direct performance comparison impractical.  Due to the lack of
recent updates to the codebase, we did not benchmark against Pydgin.

\PARAGRAPH{GenSim from the University of Edinburgh}
GenSim~\cite{WagstaffH:frohigladtfiss,WagstaffH:earpareiajcrissgfahlad,WagstaffH:autisabcaatcgfriss,SpinkT:retsysldbth}
provides a retargetable and JIT-accelerated ISS framework.  The GenSim
toolchain, using the GenC specification language, and Captive dynamic
binary translator (DBT) achieve a performance speedup of ~2$\times$
over QEMU~\cite{SpinkT:effcodgiaarbdbt}.  The performance gains are
attributed to: (i) hardware-accelerated guest memory address
translation~\cite{SpinkT:haracccafsv}; (ii) partial
evaluation~\cite{WagstaffH:earpareiajcrissgfahlad}, a technique
related to meta-tracing~\cite[Chapter 12]{bolz_meta-tracing_2014};
(iii) DBT optimised for ISS workloads~\cite{SpinkT:effcodgiaarbdbt};
(iv) parallelisation~\cite{KyleS:effparissoemcpurbjitdbt}; (v) use of
a DSL with performance-focused constructs.
We believe most of these techniques are general over the JIT/DBT system
and thus applicable to Pydrofoil as an avenue for future work.
However, while the hardware-acceleration of guest memory address translation
provides substantial performance improvement, the implementation used by
Captive is not possible in Pydrofoil. The technique requires
arranging host page tables so guest address translations can be performed
by the host's memory management unit. This use of hardware translation is
faster than the simulation of multiple levels of page table walks in
software. Unfortunately, it is only suitable
for systems where the address space can be fully managed (Captive hosts the DBT
in a bare-metal virtual machine, to gain full and unrestricted access to the host machine's
hardware); as a user-mode application Pydrofoil is incapable of that. Additionally,
this would harm one of Pydrofoil's features by no longer fully simulating the
precise model provided, and instead requiring ahead-of-time knowledge of
the guest's memory translation regime and a separate implementation for
each host-guest architecture combination.
The performance implications of compiling from Sail, compared to lower-level
languages, remain a subject of ongoing investigation. A tool was made to
translate Sail to GenC, forming a Sail frontend for GenSim, but was
unsuccessful. The internal architecture of the GenSim toolchain scales
poorly with input model size; the handwritten models are ~20,000 lines of
GenC and compile in minutes, but attempts to use Sail translated to GenC
resulted in tens of millions of lines and failed to compile even after days.
The same issue occurred when developing a tool to translate Sail into Rust;
the large volume of code being emitted caused the Rust compiler to use an
unsustainable amount of memory. This appears to be a consistent challenge
faced when working with high-level, machine-generated Sail specifications.

\section{Conclusion}\label{section_conclusion}

This paper introduces Pydrofoil, a tool designed to enhance the
performance of ISSs based on Sail. We identify the primary
bottlenecks of the current Sail system in Sections
\ref{section_background} and \ref{section_implementation}, addressing
our first research question (RQ1). We propose a staged compiler
architecture that utilises meta-tracing, as detailed in Section
\ref{section_implementation}. Our benchmarks, presented in Section
\ref{section_benchmarks}, demonstrate that Pydrofoil significantly
improves simulation speed compared to Sail. However, in its current
iteration, it does not yet reach the performance levels of a
hand-tuned system like QEMU, answering RQ2 and RQ3. Future work aimed
at closing this gap is outlined in Section \ref{sec-future}, serving
as a roadmap for achieving performance parity with QEMU.

\clearpage
\bibliography{bib/semantic}
\clearpage
\appendix

\section{Benchmark details}\label{app_benchmarks}

\subsection{SPEC}

The Standard Performance Evaluation Corporation (SPEC) benchmark
\cite{SPEChomepage} is a widely used suite of standardised tests used
to evaluate the performance of computer systems. We restrict our
attention to one key component of SPEC, the SPEC CPU benchmarks
\cite{BucekJ:specpungcb}, which measure the performance of a system's
processor and memory subsystem. SPEC CPU has two parts: SPECint and
SPECfp. The former focuses on integer operations, the latter on
floating-point operations.  Below is a brief summary of each of the
SPEC benchmarks we use: 
\begin{itemize}

\item \CODE{600.perlbench\_s}. This benchmark is based on the Perl
  programming language interpreter. It runs a series of Perl scripts
  that perform various tasks, such as manipulating strings, processing
  regular expressions, and executing system commands.

\item \CODE{602.gcc\_s}. This benchmark uses the GNU Compiler
  Collection (GCC) to compile several different source code files. It
  measures the time taken to compile these files into executable code.

\item \CODE{605.mcf\_s}. This benchmark solves a large-scale
  transportation problem using a network flow algorithm. It involves
  optimising the flow of goods through a network of nodes and edges.

\item \CODE{620.omnetpp\_s}. This benchmark simulates a large-scale
  network using the OMNeT++ discrete event simulation framework. It
  models the behaviour of network protocols and devices.

\item \CODE{623.xalancbmk\_s}. This benchmark uses the Xalan-C++
  library to transform XML documents into HTML or other formats using
  XSLT stylesheets.

\item \CODE{625.x264\_s}. This benchmark encodes a video stream using
  the x264 video encoder, which is a popular open-source
  implementation of the H.264/MPEG-4 AVC video compression standard.

\item \CODE{631.deepsjeng\_s}. This benchmark runs the DeepSjeng chess
  engine, which plays a game of chess at a high level of skill. It
  involves evaluating many possible moves and counter-moves.

\item \CODE{641.leela\_s}. This benchmark runs the Leela Chess Zero
  engine, which uses neural networks to play chess. It involves both
  traditional chess algorithms and machine learning techniques.

\item \CODE{648.exchange2\_s}. This benchmark simulates a financial
  exchange where multiple traders buy and sell assets. It involves
  complex financial calculations and decision-making.

\item \CODE{657.xz\_s}. This benchmark compresses and decompresses
  data using the XZ compression algorithm, which is based on the LZMA2
  compression method.

\end{itemize}

Each of these benchmarks is designed to stress different aspects of a
CPU's performance, from integer and floating-point calculations to
memory access patterns and complex algorithms. They are used to
evaluate how well a CPU can handle real-world tasks, making them
valuable tools for comparing different processors and architectures.

\subsection{CHERIoT's allocator-benchmark}

The ``allocator-benchmark'' can be found in
\cite{CHERIoTPlatform:alloc_bench}.  For each power of 2 from 32 to
131072, the benchmark tries to allocate 1 MB with \CODE{malloc}, and
then deallocates it using \CODE{free}. It also prints (to the emulated
character device) the time it took to allocate and deallocate each 1 MB.

\end{document}